\documentclass[doublespacing]{elsart}
\usepackage{graphicx}
\usepackage[latin1]{inputenc}
\usepackage{amssymb}
\usepackage{bm}

\begin{document}

\begin{frontmatter}

\title{Classical Dissipation and Asymptotic Equilibrium via
Interaction with Chaotic Systems}

\author{M.V.S. Bonan\c{c}a}
\and
\author{M.A.M. de Aguiar\corauthref{a1}}
\ead{aguiar@ifi.unicamp.br}
\corauth[a1]{M.A.M. de Aguiar}

\address{Instituto de F\'isica 'Gleb Wataghin',
Universidade Estadual de Campinas, \\
Caixa Postal 6165, 13083-970 Campinas, São Paulo, Brazil}

\begin{abstract}

We study the energy flow between a one dimensional oscillator and
a chaotic system with two degrees of freedom in the weak coupling
limit. The oscillator's observables are averaged over an initially
microcanonical ensemble of trajectories of the chaotic system,
which plays the role of an environment for the oscillator. We show
numerically that the oscillator's average energy exhibits
irreversible dynamics and `thermal' equilibrium at long times. We
use linear response theory to describe the dynamics at short times
and we derive a condition for the absorption or dissipation of
energy by the oscillator from the chaotic system. The equilibrium
properties at long times, including the average equilibrium
energies and the energy distributions, are explained with the help
of statistical arguments. We also check that the concept of
temperature defined in terms of the `volume entropy' agrees very
well with these energy distributions.

\end{abstract}

\end{frontmatter}

\section{Introduction}

Low dimensional chaotic systems can, under appropriate
circumstances, play the role of thermodynamical heat baths
\cite{wilkinson,berry,tulio,jarzynski,cohen99,fishman,cohen}. If a
slow system with few degrees of freedom is weakly coupled to a
fast chaotic system, the slow system's average trajectory can
dissipate energy into the chaotic one at short times.

One of the initial motivations for the consideration of low
dimensional chaotic dynamics as environments for macroscopic
systems was the work by Brown, Ott and Grebogi \cite{ott} on the
ergodic adiabatic invariant. For a chaotic Hamiltonian system with
a slowly varying parameter, the volume of the energy shell is the
only adiabatic invariant. Brown et al showed that the first order
correction to this invariant has a diffusive temporal behavior.
Besides, if such time varying parameter is thought of as a second
system coupled to the chaotic one, then this diffusive correction
of the adiabatic invariant would lead to a dissipative force on
the slow system \cite{wilkinson}.

This problem was again reformulated by Berry and Robbins
\cite{berry} in terms of a system of interest interacting with an
environment, or `thermal bath'. The average force acting on the
system of interest due to the coupling with the chaotic system can
be calculated in adiabatic approximation. The lowest order part of
this force was shown to be the Born-Oppenheimer force and the next
order to be a geometric magnetism type of force plus a
deterministic friction - a force proportional to the slow system's
velocity. Friction is therefore generated in the context of small
systems, without the need for the thermodynamical limit, and the
chaotic nature of the motion becomes the essential ingredient.
Indeed, typical correlation functions of chaotic systems decays
exponentially with the time, as opposed to the quasi periodic
behavior observed in regular systems. Jarzynski \cite{jarzynski}
showed later that coupling to a low-dimensional chaotic motion can
also lead to a 'thermalization' of the system of interest, very
much like the thermalization of a Brownian particle interacting
with a large thermal bath. Finally, Carvalho and de Aguiar
\cite{tulio} established a connection between de formalism
developed by Caldeira and Leggett \cite{caldeira} for describing
dissipation via coupling with a thermal bath and via coupling with
a chaotic system.

In this paper we revisit this problem from the classical point of
view and consider specific examples of two-dimensional chaotic
systems coupled to a one-dimensional oscillator. Our main purpose
here is to understand the energy flow between the system of
interest, which we shall call `the oscillator' and the chaotic
system to which this oscillator is coupled. We study the energy
flow at short times and the approach to equilibrium at long times.
In order to consider the chaotic system as playing the role of an
environment, we assume that the only information available from
this system is its initial energy. For the oscillator this implies
that only microcanonical averages of its observables (over the
chaotic system variables) are accessible. Therefore, a typical
numerical calculation corresponds to fix an initial condition for
the oscillator and to evolve an ensemble of trajectories whose
initial conditions for the chaotic variables are randomly selected
at a fixed (chaotic system) energy shell.

For short times our numerical simulations show that the average
energy of the oscillator may increase or decrease, absorbing
energy from the chaotic system or dissipating energy into the
chaotic system. The initial energies of both systems is what
ultimately dictates which of the two possibilities actually
occurs. In particular, there exists initial values of these
energies such that no exchange occurs on the average. We use
Linear Response Theory to study the energy flow in the short time
limit. We show that the average motion of the oscillator follows a
Langevin type of equation with frequency-dependent friction and a
quadratic correction to the oscillator potential, similar to the
Born-Oppenheimer force that appears in the adiabatic theory. We
also derive a simple condition for dissipation or absorption of
energy by the oscillator involving the ratio of the initial
energies of the systems. This theoretical prediction agrees well
with our numerical calculations for short times, but it is not
accurate to predict the long time behavior.

Our simulations show that, at long times, the average energy of
both the oscillator and the chaotic system tend to an equilibrium.
The value of these equilibrium energies depend once again on the
initial conditions. The connection between asymptotic
thermalization and initial conditions is well known for a Brownian
particle. In that case, the increase or decrease of the average
energy of the particle depends on its initial energy $E_0$ and on
the temperature $T$ of the thermal bath. The particle absorbs
energy from the reservoir if $E_0 < k_B T$ and loses energy into
it if $E_0 > k_B T$, thermalizing always at $k_B T$. Here we have
a similar situation, with the increasing or decreasing of the
average energy of the oscillator depending only on its initial
energy and on the initial energy of the chaotic system. However,
contrary to the case of the Brownian particle, the condition for
equilibrium at long times is generally different from the
condition of no energy exchange at short times. Despite the
theoretical work of Jarzynski \cite{jarzynski} on the long term
thermalization of these systems, it is still not clear how the
asymptotic states depend on the initial conditions of both
sub-systems. The energy distribution of the sub-systems at
equilibrium is also an important open issue. These distributions
are not likely to be of the Boltzmann type, since the systems are
small, and they may depend not only on the initial conditions but
also on the density of states of the systems involved
\cite{ramshaw,izrailev,costa}. In this paper we shall derive the
long time equilibrium conditions and energy distributions
explicitly for two model systems. Finally, we use the definition
of temperature proposed in \cite{adib} for small systems and check
that it agrees completely with the statistical calculation in
terms of the density of states and energy distributions.

We emphasize that our approach uses the Linear Response Theory,
and no explicit assumptions on adiabatic properties of the
oscillator is required. Despite the difficulties involved in the
calculation of the response function for microcanonical ensembles,
the formulation of this problem in terms of Linear Response Theory
is of great interest for the study of its quantum analog
\cite{cohen}. We recall that the usual formulation of quantum
dissipation \cite{caldeira} involves response functions and that
the adiabatic approximation leads to frustrating results in the
quantum formulation \cite{berry}. The present work has some
similarities with that of ref.\cite{tulio}, which also considered
the model of an oscillator coupled to a chaotic bath to study
dissipation at short times. Here we study both the short and long
time limits, showing that the coupling with the chaotic system may
indeed lead to an equilibration very much like that caused by the
coupling with a large thermal reservoir.

The outline of the paper is as follows: in section II we describe
our model systems and compute the correlation functions that are
relevant for the calculation of the response functions. In Sec.
III we review the classical theory of linear response and in Sec.
IV we apply the theory to our model. We calculate the response
functions explicitly in the microcanonical ensemble. In Sec. V we
compute the average energy and average equation of motion of the
oscillator. We also derive a condition for the initial increase or
decrease of the average energy of the oscillator for short times,
and compare it with numerical results. In Sec. VI we calculate the
long time equilibrium values of the average energies using
statistical arguments and compare them with numerical values.
Finally in section VII we summarize our conclusions.

\section{The Model and First Numerical Results}

Our model Hamiltonian consists of a one-dimensional harmonic
oscillator $H_o(z)$ coupled to a chaotic system with two degrees
of freedom $H_c(x,y)$:
\begin{equation}
\label{modelo}
H = H_o(z) + H_c(x,y) + V_I(x,z),\label{eq1}
\end{equation}
where
\begin{equation}
H_o(z) = \frac{p_z^2}{2m} + \frac{m\omega_0^2 z^2}{2},
\label{harmosc}
\end{equation}
and
\begin{equation}
V_I(x,z) = \gamma xz. \label{perturb}
\end{equation}

For the chaotic Hamiltonian we consider two systems:
\begin{eqnarray}
H_c(x,y) = \frac{p_x^2}{2} + \frac{p_y^2}{2} + (y - \frac{x^2}
{2})^2 + 0.1\frac{x^2}{2} \label{nelson}
\end{eqnarray}
and
\begin{eqnarray}
H_c(x,y) = \frac{p_x^2}{2} + \frac{p_y^2}{2} + \frac{x^2 y^2}{2} +
a\frac{(x^4+y^4)}{4} \label{quart} \;.
\end{eqnarray}
The Hamiltonian (\ref{nelson}) is known as the Nelson system (NS)
\cite{baranger} and we shall refer to (\ref{quart}) as the Quartic
system (QS) \cite{percival}. The NS exhibits soft chaos and is
fairly regular for $E_c \,_{\sim}^<\,0.05$, strongly chaotic for
$E_c \,_{\sim}^>\, 0.3$ and mixed for intermediate values of the
energy. The QS is invariant under a scaling transformation of the
coordinates and the time, which implies that the dynamics on all
energy shells are equivalent to each other. We shall explore this
property later. The QS is integrable for $a=1.0$, strongly chaotic
for $a \,_{\sim}^<\,0.1$ and mixed for intermediate values of the
parameter.

We want to investigate the situation in which the chaotic system
plays the role of an external environment for the oscillator.
Therefore, we assume that detailed information about the chaotic
system is not available. If the environment were modelled by a
heat bath, the only macroscopic relevant information would be its
temperature. In the present case we assume that the only
information available is the initial energy of the chaotic system.
For the oscillator this implies that only microcanonical averages
of its observables (over the chaotic system variables) are
accessible.

In order to characterize the two chaotic systems
Eqs.(\ref{nelson}) and (\ref{quart}), we show in Fig. 1 the NS
correlation functions $\langle x(0)x(t)\rangle_e$ and $\langle
p_x(0)x(t)\rangle_e$, for $E_c=0.38$ and, in Fig. 2, the QS
correlation functions for $a=0.1$ and $E_c=5.0$. These functions
play important roles in the linear response theory to be developed
later. The brackets $\langle.\rangle_e$ stand for the average on
the microcanonical ensemble of initial conditions. The correlation
functions are obtained numerically with a fixed time step
symplectic integration algorithm \cite{forest} applied to the
isolated chaotic system. In our calculations we used time steps of
the order of $10^{-4}$ for figures 1, 2, 5, 6 and 7 (short times)
and $10^{-2}$ for figures 3 and 4. The number of initial
conditions in each case is indicated on the figure captions.

These correlation functions can be well fitted by the expressions
\begin{equation}
\begin{array}{ll}
\langle x(0)x(t)\rangle_e &=\sigma e^{-\alpha t} \cos{\omega t},\\
\langle p_x (0) x(t)\rangle_e &=\mu e^{-\beta t} \sin{\Omega t},\\
\langle p_x (0) p_x(t)\rangle_e &=\kappa e^{-\lambda t}\cos{\nu t}
\label{correl}
\end{array}
\end{equation}
(see Figs.1 and 2), which have the proper parity of the
corresponding correlation functions. For the NS the fitting
furnishes the decay rates $\alpha=0.0418$ and $\beta=0.0456$, the
amplitudes $\sigma=1.865$ and $\mu=0.409$ and the frequencies of
oscillation $\omega=0.1963$ and $\Omega=0.2043$ with
$\chi^2\,\sim\,10^{-4}$. For the QS we obtain $ \alpha = 0.207$,
$\beta=0.208$ and $\lambda=0.206$; $\sigma=2.268$, $\mu=3.67$ and
$\kappa=4.10$; and $ \omega=1.1027 $, $\Omega=1.1481$  and
$\nu=1.189$ with $\chi^2\,\sim\,10^{-3}$. This type of correlation
function is typical of chaotic systems in general
\cite{berry,russos} and also occurs in non-isolated systems
subjected to random noise, such as a Brownian particle. The source
of the correlation loss is, however, different in each case.
Notice that the numerically calculated exponents $\alpha$, $\beta$
and $\lambda$ are all very similar and so are the frequencies
$\omega$, $\Omega$ and $\nu$.

When the coupling between the chaotic system and the oscillator is
turned on, the overall conserved energy flows from one system to the
other. The oscillator's energy, in particular, fluctuates as a
function of the time for each specific trajectory. The oscillator's
average energy is calculated by taking an ensemble of initial
conditions uniformly distributed over the chaotic energy surface
$E_c$. For the oscillator we fix only one initial condition, which we
choose to be $z(0) = 0$ and $p_z(0) = \sqrt{2mE_o (0)}$. The
microcanonical ensemble of initial conditions is propagated and, at
each instant, $H_o$ is calculated for each trajectory and its average
value is computed. Figs. 3 and 4 show the oscillator's average energy
$\langle E_o (t)\rangle$ as a function of the time for different
values of $E_o (0)$. We observe that the average effect of the
interaction with the chaotic system leads to a 'thermalization' of
$\langle E_o(t)\rangle$. We also see that $E_o (0)$ plays an important
role on the long term behavior of $\langle E_o (t)\rangle$. As we
shall see later using linear response theory, the ratio $E_o (0)/ E_c
(0)$ defines the initial rate of energy variation in time. Zooming the
short time behavior of the curves in Figs. 3 and 4 (see Figs.  5 and
6) shows fast oscillations of $\langle E_o (t)\rangle$. As we shall
see, part of these oscillations are due to a change in the effective
potential acting of the oscillator, similar to the Born-Oppenheimer
force of the adiabatic theory (see eq.(\ref{eor}) in section
5). Notice that the oscillator's equilibrium energy is different for
each situation, which clearly distinguishes the small chaotic
environment from an infinite thermal bath.

As a last remark we note that, since the environment has few
degrees of freedom, the oscillator's motion for a single
realization (single initial condition) exhibits large fluctuations
with respect to the average.

\section{Linear response theory}

The calculation of averages such as $\langle E_o (t)\rangle$
involve the calculation of the microcanonical distribution
function $\rho(q,p;t)$ whose initial value is $\rho(q,p;0) =
\delta(H_c(q,p)-E_c(0))/\Sigma(E_c(0))$, with $\Sigma(E_c(0))=\int
\mathrm{d}q\mathrm{d}p\, \delta (H_c(q,p)-E_c(0))$. If the chaotic
system were isolated, $\rho$ would be an invariant distribution
and $\rho(q,p;t)=\rho(q,p;0)$. The coupling, however, causes the
value of $H_c(q(t),p(t))$ to fluctuate in time, distorting the
energy surface $H_c=E_c(0)$ \cite{ott2}. Linear response theory
provides a way to calculate the first order corrections to this
distribution in the limit of weak coupling \cite{kubo,fick}.

Consider a Hamiltonian $H(q,p)$ perturbed by a term of the form
\begin{equation}
H_1=-A(q,p) X(t)\label{pertur} \, ,
\end{equation}
where $A$ is an arbitrary function of the coordinates and momenta
and $X(t)$ is a function of the time. A generic distribution
function $\rho(q,p)$ follows the Liouville equation
\begin{equation}
\frac{\partial\rho(t)}{\partial t}=i[L+L_1(t)]\rho(t)\label{liouville},
\end{equation}
where the Liouville operators are given by
\begin{eqnarray}
iL\rho=\sum\left(\frac{\partial H}{\partial q}\frac{\partial
\rho}{\partial p}-\frac{\partial H}{\partial p}\frac{\partial
\rho}{\partial q}\right)  =\{H,\rho\},
  \nonumber \\
iL_1\rho=\sum\left(\frac{\partial H_1(t)}{\partial
q}\frac{\partial \rho }{\partial p}-\frac{\partial
H_1(t)}{\partial p}\frac{\partial \rho}{\partial q}\right) =
\{H_1(t),\rho\},
\end{eqnarray}
where the summation is taken over the whole set of canonical
variables.

The differential equation (\ref{liouville}) has an integral
representation of the form
\begin{eqnarray}
\rho(t)=e^{i(t-t_0)L}\rho(t_0)+\int_{t_0}^t\mathrm{d}s\,
e^{i(t-s)L}iL_1(s)\rho(s)\label{intliou}.
\end{eqnarray}

Even if $\rho(t_0)$ is an equilibrium distribution of $H$
(canonical or microcanonical, for example), it will not remain
invariant if the perturbation $H_1$ is present. However, if the
perturbation is small, we may expand Eq.(\ref{intliou}) to first
order in $iL_1$ to obtain
\begin{eqnarray}
\rho(q,p;t)=\rho_e(q,p)+\int_{t_0}^t\mathrm{d}s\,
e^{i(t-s)L}\{H_1(s),\rho_e(q,p)\}\label{resplin},
\end{eqnarray}
where $\rho_e$ denotes the initial equilibrium distribution and
$\rho(q,p;t)$ is the time dependent non-equilibrium distribution
up to first order in the perturbation $H_1(t)$. Within this
approximation, the average value of a general function $B(q,p)$
can be computed as
\begin{eqnarray}
\langle B\rangle(t) =
\int\mathrm{d}q\mathrm{d}p\,\rho_e(q,p)B(q,p) +
\int_{t_0}^t\mathrm{d}s\, \phi_{BA}(t-s) X(s) \label{bmedio},
\end{eqnarray}
where the {\it response function} $\phi_{BA}(t)$ is given by
\cite{kubo}
\begin{eqnarray}
\phi_{BA}(t) &=& \int\mathrm{d}q\mathrm{d}p\,B(q,p)e^{itL} \,
\{-A(q,p),\rho_e(q,p)\} \nonumber \\
&=&\int\mathrm{d}q\mathrm{d}p\,\{\rho_e(q,p),A(q,p)\}B(q(t),p(t)),
\end{eqnarray}
with $q(t)=e^{itL}q$ and $p(t)=e^{itL}p$. Finally, expanding the
Poisson brackets and integrating by parts we obtain
\begin{eqnarray}
\phi_{BA}(t)&=&\int\mathrm{d}q\mathrm{d}p\,
\rho_e(q,p)\{A(q,p),B(q(t),p(t))\} \nonumber \\
&=&\langle\{A(q,p),B(q(t),p(t))\}\rangle_e.
\label{response}
\end{eqnarray}

The application of this theory to our model is as follows: we
consider $V_I$ as a perturbation on $H_c$, the coordinate $z$ in
$V_I$ being the analog of $X(t)$ in (\ref{pertur}). The initial
microcanonical distribution for the chaotic system is then
propagated according to Eq.(\ref{resplin}). Averages of
observables involving the chaotic system variables can also be
calculated with the help of Eq.(\ref{bmedio}). In addition to the
consideration that the oscillator acts as a perturbation to the
chaotic system, we also consider the response of the chaotic
system as a perturbation to the oscillator. This allows for the
perturbative calculation of the oscillator's average trajectory
and energy as well.

\section{Application to the model}

According to Eqs.(\ref{modelo})-(\ref{perturb}), the equation of
motion for the oscillator is given by
\begin{equation}
\ddot{z}(t) + \omega_0^2 z(t) = -\frac{\gamma}{m}x(t) \,.
\label{zoft}
\end{equation}
This equation can be solved explicitly in the form
\begin{eqnarray}
\label{osceq} &z(t)=z_d (t)-\frac{\gamma}{m}\int_0^t
\mathrm{d}s\Gamma(t-s)x(s)
\end{eqnarray}
where $z_d$ is the decoupled solution and $\Gamma(t-s) =
\sin{[\omega_0(t-s)]}/\omega_0$.

Differentiating (\ref{osceq}) with respect to $t$ and multiplying
by $m$ we obtain
\begin{eqnarray}
&p_z (t) = p_{z_d} (t)-\gamma\int_0^t \mathrm{d}s\chi(t-s)x(s)
\end{eqnarray}
with $\chi(t-s) = \cos{[\omega_0 (t-s)]}$. Taking $z(0)=0$ the
decoupled solutions $z_d (t)$ and $p_{z_d} (t)$ become
\begin{equation}
z_d (t)=\frac {p_z (0)}{m\omega_0}\sin{(\omega_0t)} \qquad p_{z_d}
(t)=p_z (0)\cos{(\omega_0t)}
\end{equation}
with $p_z (0)=\sqrt{2m E_o (0)}$. The oscillator's average energy
$\langle E_o (t)\rangle$ can be computed as
\begin{equation}
\langle E_o (t)\rangle=\frac{\langle p_z^2 (t)\rangle}{2m}+\frac{m
\omega_0^2 \langle z^2 (t)\rangle}{2} \label{enerav}
\end{equation}
with
\begin{eqnarray}
\label{p2av}
\langle p_z^2 (t)\rangle = p^2_{z_d}(t)-2\gamma
p_{z_d} (t)\int_0^t
\mathrm{d}s\chi(t-s)\langle x(s)\rangle+\nonumber \\
\gamma^2\int_0^t\mathrm{d}s\int_0^t\mathrm{d}u
\chi(t-s)\chi(t-u)\langle x(s)x(u)\rangle\\ \nonumber
\end{eqnarray}
and
\begin{eqnarray}
\label{x2av} \langle z^2 (t)\rangle=z^2_{d}(t)-\frac{2\gamma}{m}
z_d (t)\int_0^t
\mathrm{d}s\Gamma(t-s)\langle x(s)\rangle+\nonumber \\
\frac{\gamma^2}{m^2}\int_0^t\mathrm{d}s\int_0^t\mathrm{d}u \Gamma(t-s)
\Gamma(t-u)\langle x(s)x(u)\rangle.
\end{eqnarray}
Notice that the quantity $\langle E_o (t)\rangle$ is very
different from $E_o(\langle z \rangle, \langle p_z \rangle)$,
which is the energy of the average trajectory, considered in
\cite{tulio}.

Linear response theory now furnishes
\begin{equation}
\label{xave} \langle x(t)\rangle=\langle x(t)\rangle_e -
\gamma\int_0^t\mathrm{d}s \phi_{xx} (t-s)z(s)
\end{equation}
where $\phi_{xx} (t) = \langle\{x(0),x(t)\}\rangle_e$ is the
response function given by Eq.(\ref{response}) with $A=B=x$. Since
$H_c(-x)=H_c(x)$ for both chaotic systems given by
Eqs.(\ref{nelson}) and (\ref{quart}), $\langle x(t)\rangle_e=0$.
Substituting (\ref{xave}) into (\ref{p2av}) and (\ref{x2av}) and
considering terms up to order $\gamma^2$ we obtain
\begin{equation}
\begin{array}{ll}
\langle p^2_z (t)\rangle &=
p^2_{z_d}(t)+2\gamma^2\int_0^t\mathrm{d}s\chi(t-s)
\int_0^s\mathrm{d}u\phi_{xx}(s-u)z_d(u)  \\
 & + \gamma^2\int_0^t \mathrm{d}s\int_0^t
\mathrm{d}u\chi(t-s)\chi(t-u) \langle x(s)x(u)\rangle_e,
\label{pz2}
\end{array}
\end{equation}
\begin{equation}
\begin{array}{ll}
\langle z^2(t)\rangle & =
z^2_d(t)+\frac{2\gamma^2}{m^2}\int_0^t\mathrm{d}s
    \Gamma(t-s)\int_0^s\mathrm{d}u\phi_{xx}(s-u)z_d(u) \\
&+\frac{\gamma^2}{m^2}\int_0^t \mathrm{d}s \int_0^t
  \mathrm{d}u\Gamma(t-s)\Gamma(t-u)\langle x(s)x(u) \rangle_e.
\label{z2}
\end{array}
\end{equation}

Eqs. (\ref{pz2}) and (\ref{z2}) represent the average behavior of
the harmonic oscillator when interacting with a generic
x-symmetric system through the perturbation $V_I$. They show that
all the influence of the interacting system is contained in the
functions $\langle x(0)x(t)\rangle_e$ and $\phi_{xx}(t)$. If the
interacting system is integrable, these functions exhibit
quasi-periodic oscillations. For chaotic systems, on the other
hand, they typically decay exponentially with time, leading to
qualitatively distinct average results.

The response function is given by $\phi_{xx} (t) =
\langle\{x(0),x(t)\}\rangle_e$ where \{.\} is the Poisson bracket
with respect to the initial conditions \cite{kubo} . Since
$p_x(0)$, $p_y(0)$, $x(0)$ and $y(0)$ are independent variables we
integrate by parts and obtain
\begin{eqnarray}
\phi_{xx} (t) = \int\mathrm{d}V\{\rho_e,x(0)\}x(t) =
-\int\mathrm{d}V \frac{\partial\rho_e}{\partial p_x(0)} x(t)
\end{eqnarray}
where  $\mathrm{d}V = \mathrm{d}x(0) \mathrm{d}y(0) \mathrm{d}
p_x(0)\mathrm{d}p_y(0)$,
\begin{equation}
\rho_e = \delta(H_c-E_c(0))/\Sigma(E_c(0)) \label{rhoeq}
\end{equation}
is the microcanonical distribution and
\begin{equation}
\Sigma(E) = \int \delta(H_c-E) \, \mathrm{d}V \label{rhonorm}
\end{equation}
is a normalization factor. We obtain
\begin{equation}
\phi_{xx}(t) =
-\frac{1}{\Sigma(E_c(0))}\int\mathrm{d}V\frac{\partial
\delta(H_c-E_c(0))}{\partial H_c}p_x(0)x(t). \label{phi1}
\end{equation}
This integral can be performed explicitly by changing to variables
across the energy shell, $H_c$, and along the energy shell, $\xi$,
$\theta$ and $\varphi$:
\begin{equation}
\begin{array}{ll}
\phi_{xx}(t) &= \displaystyle{\int\mathrm{d}H_c\mathrm{d}\Theta
 \frac{\delta(H_c-E_c(0))}{\Sigma(E_c(0))}
 \frac{\partial}{\partial H_c}
\left( J p_x(0)x(t)\right)} \\  \\
 =&\left\langle p_x(0)x(t) \frac{\partial J/\partial H_c}{J} \right\rangle_e +
\left\langle\frac{\partial}{\partial H_c} (p_x (0)x(t))
\right\rangle_e  \label{phi2}
\end{array}
\end{equation}
where $\Theta = (\xi, \theta, \varphi)$,
$J=J(H_c,\xi,\theta,\varphi)$ is the Jacobian of the
transformation and the derivatives are computed at $E_c(0)$.
Similar results for response functions calculated on the
microcanonical ensemble can found in \cite{bia94}. For the NS, the
explicit transformation is given by
\begin{equation}
\begin{array}{ll}
x=\left(\frac{1}{0.05}\right)^{1/2}\sqrt{H_c}\cos{\xi}\cos{\theta}
& \quad y=\frac{x^2}{2}+\sqrt{H_c}\sin{\xi}\cos{\varphi} \nonumber \\
p_x=\sqrt{2 H_c}\cos{\xi}\sin{\theta}
 & \quad p_y=\sqrt{2 H_c}\sin{\xi}\sin{\varphi}
\end{array}
\end{equation}
with $J=H_c f_N(\Theta)$. The response function simplifies to:
\begin{equation}
\phi_{xx}(t) = \frac{\langle p_x(0)x(t)\rangle_e}{E_c (0)} +
\left\langle\frac{\partial}{\partial H_c} (p_x (0)x(t))
\right\rangle_e \;.
\end{equation}
For $H_c$ close to $E_c(0)$ we assume that the amplitude of the
correlation function $\langle p_x(0)x(t)\rangle_e$ varies linearly
with $H_c$ and we make the approximation
\begin{equation}
\left\langle\frac{\partial}{\partial H_c} (p_x(0)x(t))
\right\rangle_e \approx \frac{\langle p_x(0)x(t)\rangle_e}{E_c(0)}
\,. \label{phiaprox1}
\end{equation}
We obtain
\begin{equation}
\phi_{xx}(t)=\frac{2}{E_c(0)}\langle p_x(0)x(t)\rangle_e,
\end{equation}
which is similar to the canonical response function \cite{kubo}.

For QS, the explicit transformation is given by
\begin{equation}
\begin{array}{ll}
x^2 &= \sqrt{\frac{2 H_c}{\cos{2\theta}}}
\left[\frac{\cos{\theta}}{\sqrt{1+a}}
+\frac{\sin{\theta}}{\sqrt{1-a}}\right] \sin{\xi} \\
y^2 &= \sqrt{\frac{2 H_c}{\cos{2\theta}}}
\left[\frac{\cos{\theta}}{\sqrt{1+a}}
-\frac{\sin{\theta}}{\sqrt{1-a}}\right] \sin{\xi}\\
p_x &=\sqrt{2 H_c}\cos{\varphi}\cos{\xi} \\
p_y &=\sqrt{2 H_c}\sin{\varphi}\cos{\xi}
\end{array}
\end{equation}
with $0 < \theta < \theta_0 < \pi/4$ and
$\tan{\theta_0}=\sqrt{(1-a)/(1+a)}$. The jacobian is given
$J=H^{1/2}_c f_Q(\Theta)$ and the response function (\ref{phi2})
simplifies to:
\begin{eqnarray}
\phi_{xx}(t)=\frac{\langle p_x(0)x(t)\rangle_e}{2 E_c (0)} +
\left\langle\frac{\partial}{\partial H_c} (p_x (0)x(t))
\right\rangle_e.  \label{phi3}
\end{eqnarray}

The last term of (\ref{phi3}) can be calculated explicitly because
the QS has an energy scaling dynamics \cite{percival} given by:
\begin{eqnarray}
p_x(t)=\left(\frac{E_c}{E_c'}\right)^{1/2}p'_x(t'), \quad
x(t)=\left( \frac{E_c}{E_c'}\right)^{1/4}x'(t'), \quad
t=\left(\frac{E_c}{E_c'}\right)^{-1/4}t'. \label{t1}
\end{eqnarray}
where $p'_x$, $x'$ and $t'$ represent the motion with the energy
$E_c'$ and $p_x$, $x$ and $t$ the motion with energy $E_c$.

Using (\ref{t1}) in (\ref{phi3}), we obtain
\begin{eqnarray}
&\frac{\partial}{\partial
H_c}(p_x(0)x(t))=\frac{\partial}{\partial H_c}
\left[\left(\frac{H_c}{E_c(0)}\right)^{3/4}p'_x(0)x'(t')\right]&
\nonumber \\
&=\frac{3}{4}\frac{1}{E^{3/4}_c(0)H^{1/4}_c}p'_x(0)
x'(t')+\left(\frac{H_c} {E_c(0)}\right)^{3/4}p'_x(0)\frac{\partial
x'(t')}{\partial H_c}&
\end{eqnarray}
where the last term is given by
\begin{eqnarray}
&\left(\frac{H_c}{E_c(0)}\right)^{3/4}p'_x(0)\frac{\partial x'(t')}
{\partial H_c}=\left(\frac{H_c}{E_c(0)}\right)^{3/4}p'_x(0)
\frac{dx'(t')}{dt'}\frac{dt'}{dH_c}=& \nonumber \\
&=\left(\frac{H_c}{E_c(0)}\right)^{3/4}\frac{p'_x(0)p'_x(t')}{4}
\frac{t}{E^{1/4}_c(0)H^{3/4}_c}=\frac{t\,p'_x(0)p'_x(t')}{4E_c(0)}.&
\end{eqnarray}
Thus,
\begin{eqnarray}
\left\langle\frac{\partial}{\partial
H_c}(p_x(0)x(t))\right\rangle_e= \frac{3\langle
p_x(0)x(t)\rangle_e}{4 E_c(0)}+ \frac{t\langle
p_x(0)p_x(t)\rangle_e}{4 E_c(0)}\label{aprox2}
\end{eqnarray}
and, from (\ref{phi3}) and (\ref{aprox2}),
\begin{eqnarray}
\phi_{xx}(t)=\frac{5}{4}\frac{\langle
p_x(0)x(t)\rangle_e}{E_c(0)}+ \frac{t\langle
p_x(0)p_x(t)\rangle_e}{4 E_c(0)}.
\end{eqnarray}
The simplification we obtained, due to the specific form of the
Jacobian, is valid for a large class of Hamiltonians, including
the billiards.

\section{Short time dynamics}

In order to derive some explicit formulas, we consider the
expressions (\ref{correl}) for the correlation functions derived
in section 2 by fitting the numerical results. For the NS we
assume $\Omega\approx\omega$ and $\beta\approx\alpha$. Using Eqs.
(\ref{enerav}), (\ref{pz2}) and (\ref{z2}) we obtain the following
result:
\begin{equation}
\langle E_o(t)\rangle = E_o(0)+\frac{\gamma^2}{m}(B+At+f(t)+g(t)),
\label{eoft}
\end{equation}
where $B$ is a constant, $f(t)$ is an oscillatory function and
$g(t)$ is proportional to $e^{-\alpha t}$. The important result is
the coefficient $A$
\begin{equation}
A=4\mu\omega\alpha \frac{\left[\frac{\sigma}{4\mu\omega}
(\omega^2_0+\omega^2+\alpha^2)-\frac{E_o(0)}
{E_c(0)}\right]}{[(\omega_0-\omega)^2+\alpha^2][(\omega_0+
\omega)^2+\alpha^2]}. \label{coefa}
\end{equation}
For fixed oscillator frequency $\omega_0$ and a given chaotic
energy shell $E_c(0)$ (and, consequently, for given $\sigma$,
$\mu$, $\omega$ and $\alpha$), the ratio $E_o(0)/E_c(0)$ is the
responsible for the average increase or decrease of $\langle
E_o(t)\rangle$. The short time equilibrium in the energy flow is
given by the condition $A=0$, or
\begin{equation}
\frac{E_o(0)}{E_c(0)}=\frac{\sigma}{4\mu\omega}(\omega^2_0+\omega^2
+\alpha^2) \;. \label{equil}
\end{equation}

We now turn to the equation of motion of $z(t)$ under the average
effect of the interaction with the chaotic system. From
Eqs.(\ref{zoft}) and (\ref{xave})
\begin{equation}
\langle\ddot{z}(t)\rangle+\omega^2_0 \langle z(t)\rangle
=-\frac{\gamma}{m}\langle x(t)\rangle
  =\frac{\gamma^2}{m}\int_0^t\mathrm{d}s \, \phi_{xx} (t-s)
  \langle z(s)\rangle.
\end{equation}

Integrating by parts yields
\begin{eqnarray}
\langle\ddot{z}(t)\rangle+\left(\omega^2_0-\frac{\gamma^2
F(0)}{m}\right) \langle z(t)\rangle+
\frac{\gamma^2}{m}\int_0^t\mathrm{d}s \, F(t-s) \langle
 \dot{z}(s)\rangle+ \frac{\gamma^2}{m}z(0)F(t)=0 \label{zaveoft}
\end{eqnarray}
where
\begin{equation}
F(t-s) =\frac{2\mu
e^{-\alpha(t-s)}}{E_c(0)(\alpha^2+\omega^2)}\{\omega
\cos{[\omega(t-s)]} +\alpha\sin{[\omega(t-s)]}\}. \label{fts}
\end{equation}

Eq.(\ref{zaveoft}) shows that the interaction produces a harmonic
correction to the original potential, a frequency-dependent
dissipative term and an external force proportional to $z(0)$. The
choice $z(0)=0$ simplifies (\ref{zaveoft}) and turns it into an
average Langevin equation.

We note that the correction to the harmonic potential is repulsive
and may qualitatively change the average motion of the oscillator
if $\gamma^2 > m\omega_0^2/F(0)$. The critical perturbation
$\gamma_c^2 = m\omega_0^2/F(0)$ is related to the stability of the
equilibrium points of the full three-dimensional system, which
becomes unstable for large $\gamma$, establishing a direction of
unbounded motion. For the NS, for example, the origin
$(x,y,z)=(0,0,0)$ becomes unstable at $\gamma^2 =0.1 m/\omega_0^2$
(see Eq.(\ref{nelson})). The function $F(0)$, on the other hand,
can be computed numerically from Eq.(\ref{fts}) and results in
approximately $10$. This value can also be estimated by noticing
that $F(t)$ has dimensions of $\langle x(0)x(t)\rangle/E_c$,
which, at $t=0$, yields $2\langle x(0)^2\rangle/E_c = 1/0.1$.
Therefore one must choose $\gamma < \gamma_c$ in order to consider
the coupling as a perturbation to the original system, avoiding
the introduction of these instabilities. We remark that similar
corrections are observed in the theory of Caldeira and Leggett
\cite{caldeira} and in the adiabatic calculations of
Wilkinson-Berry-Robins \cite{berry}.

Fig. 5 shows a comparison between the numerically calculated
`bare' oscillator energy $\langle E_o(t)\rangle$, where
\begin{equation}
E_o(z) = \frac{p_z^2}{2m} + \frac{m\omega_0^2 z^2}{2},
\end{equation}
the `re-normalized' oscillator energy $\langle E_{or}(t)\rangle$,
where
\begin{equation}
E_{or}(z) = \frac{p_z^2}{2m} + \frac{m\omega_0^2 z^2}{2}
-\frac{\gamma^2}{m}F(0)z^2, \label{eor}
\end{equation}
and the expression (\ref{eoft}) without the oscillating term
$f(t)$. We have chosen $\gamma$ and $m$ so that $\omega^2_0 -
\gamma^2 F(0)/m>0$. We also chose $\omega_0$ so that $g(t)$
decreases very fast, i.e., $e^{-\alpha/\omega_0}\approx 10^{-4}$.
In this case only the linear and the oscillating terms in
Eq.(\ref{eoft}) are important. We have subtracted the oscillating
part of Eq.(\ref{eoft}) in figure 5 to highlight the linear
increase or decrease in the average energy. In the time scale of
Fig.5, which corresponds to several periods of the decoupled
oscillator, the linear behavior describes very well the numerical
results. Fig. 5(b) shows the equilibrium situation according to
Eq.(\ref{equil}).

For the QS the correlation functions can also be simplified with
the assumption that $\beta\approx\alpha\approx\lambda$ (we shall
keep for now $\omega\not=\Omega\not=\nu$). We get from
(\ref{enerav}), (\ref{pz2}) and (\ref{z2})
\begin{equation}
\langle E_o(t)\rangle = E_o(0)+\frac{\gamma^2}{m}(B'+A't+f'(t)+g'(t)),
\label{eneravq}
\end{equation}
where $B'$ is a constant, $f'(t)$ is an oscillatory function and
$g'(t)$ is a sum of terms proportionals to $e^{-\alpha\,t}$ and
$t\,e^{-\alpha\,t}$. The coefficient $A'$ is given by
\begin{eqnarray}
A'=\frac{5\beta\Omega\alpha}{2}\frac{\left[\frac{2\Lambda\sigma}{5\beta\Omega}
(\omega^2_0+\omega^2+\alpha^2)-\eta\frac{E_o(0)}{E_c(0)}\right]}
{\left[(\omega_0-\omega)^2+\alpha^2\right]\left[(\omega_0+\omega)^2+
\alpha^2\right]}, \label{coefaq}
\end{eqnarray}
where
\begin{eqnarray}
\Lambda=\frac{\left[(\omega_0-\Omega)^2+\alpha^2\right]
\left[(\omega_0+\Omega)^2+\alpha^2\right]}{\left[(\omega_0-\omega)^2+
\alpha^2\right]\left[(\omega_0+\omega)^2+\alpha^2\right]}
\end{eqnarray}
and
\begin{eqnarray}
\eta&=&1+ \frac{\mu\left[(\omega_0-\Omega)^2+\alpha^2\right]
\left[(\omega_0+\Omega)^2+\alpha^2\right]}{5\beta\Omega}×\nonumber \\
& &\frac{\left[(\omega^2_0+\kappa^2)^2+2\nu^2(\omega_0+\kappa)(\omega_0-\kappa)
-3\nu^4\right]}{\left[(\omega_0-\nu)^2+\kappa^2\right]^2\left[(\omega_0+\nu)^2
+\kappa^2\right]^2}
\end{eqnarray}

We note that if we set $\kappa=0$ (which amounts to cancel the
last term of (\ref{aprox2})) and with $\omega=\Omega$,
(\ref{coefaq}) becomes very similar to (\ref{coefa}). For
$\kappa\not=0$, $\eta$ can be positive or negative and if
$\eta\,<0$, $\langle E_o(t)\rangle$ cannot decrease with time.
This possibility never happens for the NS. The equilibrium in the
energy flow is given by the condition:
\begin{eqnarray}
\frac{E_o(0)}{E_c(0)}=\frac{2\Lambda\sigma}{5\beta\Omega}
\frac{(\omega^2_0+\omega^2+\alpha^2)}{\eta}.\label{equilq}
\end{eqnarray}

The equation (\ref{zaveoft}) remains valid for QS, but the function
$F(t)$ is given by:
\begin{eqnarray}
F(t-s)&=&\int\mathrm{d}s \,\phi_{xx}(t-s) \nonumber \\
&=&\frac{\mu(t-s)e^{-k(t-s)}}{4E_c(0)(k^2+\nu^2)}\{k\cos[\nu(t-s)]
-\nu\sin[(\nu(t-s)]\} \nonumber \\
&-&\frac{\mu e^{-k(t-s)}}{4E_c(0)(k^2+\nu^2)^2}\{(\nu^2-k^2)\cos[\nu(t-s)]
+2\nu k\sin[\nu(t-s)]\} \nonumber \\
&+&\frac{5\beta
e^{-k(t-s)}}{4E_c(0)(\Omega^2+k^2)}\{\Omega\cos[\nu(t-s)]
+k\sin[\nu(t-s)]\}\, .
\end{eqnarray}

Figure 6 shows the comparison between the numerically calculated
$\langle E_o(t)\rangle$ and $\langle E_{or}(t)\rangle$ and the
expression (\ref{eneravq}) without $f'(t)$. The parameters were
chosen in such a way that $e^{-\alpha/\omega_0}\approx 10^{-9}$
and $e^{-\alpha/\omega_0}/ \omega_0 \approx 10^{-7}$, so that the
linear and oscillating terms are the most important. Again we have
subtracted $f'(t)$ to highlight the linear behavior and we have
chosen $\gamma$ and $m$ so that $\omega_0^2-\gamma^2F(0)/m>0$.

The agreement between numerical and analytical results for the
equilibrium condition is only reasonable for the QS. The amplitude
of the oscillations in $\langle E_o(t)\rangle$ are bigger than in
the NS case because of the different energy scale and coupling
strength ($E_c(0)=0.38$, $\gamma=0.006$ for the NS and
$E_c(0)=5.0$, $\gamma=0.01$ for the QS). In the approximation
$\alpha=\beta=\lambda$ and $\omega\not=\Omega\not=\nu$, the
equilibrium condition $A'=0$ gives $E_o(0)/E_c(0)=0.45$, but if we
use $\omega=\Omega\not=\nu$, it changes to $E_o(0)/E_c(0)=0.55$.
This shows that small changes in the frequencies of the
correlation functions change significantly the value of
$E_o(0)/E_c(0)$ for the equilibrium condition. Fig. 7 displays
$\langle E_o(t)\rangle$ for others values of $E_o(0)/E_c(0)$. It
shows that the slope of the average energy as a function of the
ratio $E_o(0)/E_c(0)$ has a very shallow minimum, which justifies
its poor determination with linear perturbation theory. Fig.7
suggests that the equilibrium condition may actually be located
between $E_o(0)/E_c(0)=2.5$ and $E_o(0)/E_c(0)=3.0$.

\section{Long time behavior and `thermal' equilibrium}

For long times the oscillator and the chaotic system reach an
equilibrium, in the sense that their average energies tend to
constant values. This is shown in Figs. 3 and 4 for the NS and the
QS respectively. Notice that the energy ratio at the asymptotic
equilibrium, at long times, is not necessary related to the
equilibrium condition at short times, since non-linear effects are
certainly important in the later. In order to characterize the
equilibrium it is important to understand the energy distribution
within each sub-system. It is particularly interesting to check if
the oscillator follows Boltzmann's distribution.

To obtain the distributions numerically, we construct histograms
in which the values of $E_o$ and $E_c$ are extracted from each
trajectory of the ensemble for a fixed (long) time. The energy
axis is divided into bins and the number of trajectories of the
ensemble for which the oscillator's energy fall into each bin is
counted. The same process is performed with respect to the chaotic
system's energy. Figure 8 shows the energy distributions thus
obtained for the oscillator and the chaotic systems, NS and QS.
The vertical axis shows the counts (occurrences) in percents of
the total number of trajectories. In both cases it is clear that
the oscillator does not follow the Boltzmann exponential law. This
is actually not surprising, since the chaotic system is small
(with only two degrees of freedom) and thus its energy is
comparable to that of the oscillator. The Boltzmann distribution
comes out naturally only when the system of interest is in contact
with a reservoir of much larger energy (spread among its many
degrees of freedom) \cite{reif}, a condition that is not fulfilled
here. In this context, it is also not clear whether a temperature
can be defined in the present situation. We shall return to this
question in a moment.

The energy distribution of the subsystems can be totally
understood in terms of their density of states. To see this, we
assume that two hypothesis that are usual in the statistical
physics of large systems \cite{reif} are also valid for our small
systems. The first is to consider that, at equilibrium, all states
of the full system, Eq.(\ref{modelo}), are equally probable, i.e.,
the hypothesis of equal probabilities {\it a priori}. The second
assumption concerns weak interactions between the subsystems. The
number of states $dN(E)$ {\em of the full system}, for which the
oscillator has energy between $E$ and $E+dE$, can be written in
terms of the density of states
\begin{equation}
n(\epsilon)=\int \mathrm{d}V \, \delta(H(x,y,z)-\epsilon),
\label{denstates}
\end{equation}
where $H(x,y,z)$ is given by Eq.(\ref{eq1}) and
$\mathrm{d}V=\mathrm{d}x \mathrm{d}p_x \mathrm{d}y \mathrm{d}p_y
\mathrm{d}z \mathrm{d}p_z$. Neglecting the interaction potential
$V_I$ in Eq.(\ref{eq1}), we can write $dN(E) = n_o(E) \,
n_c(E_T-E) \, \mathrm{d}E$, where $n_o$ and $n_c$ are the
oscillator and chaotic system densities of states, respectively,
and $E_T$ is the total energy.

Based on these two assumptions, we can calculate the probability
that the oscillator has energy between $E$ and $E+dE$. The
probability density $p_o(E)$ is, because of the first assumption,
proportional to the number of states $dN(E)$:
\begin{eqnarray}
p_o(E)dE \propto dN(E) = n_o(E)n_c(E_T-E)dE.
\end{eqnarray}
Likewise, the probability density of the chaotic system is
\begin{eqnarray}
p_c(E) dE \propto n_c(E)n_o(E_T-E) dE.
\end{eqnarray}
We find that $n_o(E)$ is a constant and that $n_c(E)\propto E$ for
NS and $n_c(E)\propto E^{1/2}$ for QS. Thus we obtain
\begin{equation}
\left\{
\begin{array}{l}
p_o(E) \propto (E_T-E) \\
p_c(E) \propto E
\end{array} \right. \qquad \qquad \qquad \mbox{for the NS}
 \label{probn}
\end{equation}
and
\begin{equation}
\left\{
\begin{array}{l}
p_o(E) \propto (E_T-E)^{1/2} \\
p_c(E) \propto E^{1/2}
\end{array} \right. \qquad \qquad \quad \mbox{for the QS}
 \label{probq}
\end{equation}
Of course these expressions are meaningful only for $0 < E < E_T$.

The full lines in Fig. 8 show a linear fitting for the case of the
NS and a square root fitting for the QS. The fittings agree very
well with the numerical results. The sudden decrease of the
numerical distributions of the chaotic systems for high energies
is due to the constraint that $E_T$ is fixed.

Finally, with Eqs.(\ref{probn}) and (\ref{probq}), we can also
calculate the oscillator's average energy in the equilibrium and
compare the results with the values in figures 3 and 4. We have
\begin{eqnarray}
\bar{E_o}=\int^{E_T}_0\mathrm{d}E\,\frac{p_o(E)}{Z}\,E,
\end{eqnarray}
where the normalization constant
\begin{eqnarray}
Z=\int_0^{E_T}\mathrm{d}E\,p_o(E)
\end{eqnarray}
is $Z=E^2_T/2$ for the NS and $Z=2E^{3/2}_T/3$ for the QS. We
obtain $\bar{E_o}=E_T/3$ for NS and $\bar{E_o}=2E_T/5$ for QS.
From the probability densities $p_c$ for N.S. and Q.S., we obtain
$\bar{E_c}=2E_T/3$ and $\bar{E_c}=3E_T/5$, respectively, for N.S.
and Q.S.. For the parameters of Fig. 3a, $E_c(0)=0.38$ and
$E_o(0)/E_c(0)=1.0$, we find $\bar{E_o}=0.253$ and for Fig. 3b,
where $E_c(0)=0.38$ and $E_o(0)/E_c(0)=0.1$, we get $\bar{E_o}=0.139$.
For Figs. 4a and 4b we obtain $\bar{E_o}=12$ and $\bar{E_o}=2.4$
respectively. In all cases the numerical values of the oscillator
equilibrium energies is very close to the statistical prediction.

We now return to the question about the temperature. We will show
that not only it is possible to define a temperature for our
subsystems but also that the equilibrium predicted by equating
these temperatures results in the same partition of average
energies predicted by the statistical analysis above. We first
consider the usual definition of temperature given by
\begin{eqnarray}
\frac{1}{T}=\frac{\partial S}{\partial E}. \label{tempdef}
\end{eqnarray}
where
\begin{eqnarray}
S=k_B\ln n(E),\label{entrop1}
\end{eqnarray}
is the entropy and $n(E)$ is the density of states given by
(\ref{denstates}). The thermal equilibrium between the oscillator
and the chaotic system implies $\partial S_o/\partial E_o =
\partial S_c/\partial E_c$. However, because $n_o(E)$ does not depend on
$E$, this gives $T_o=T_c=\infty$, and the equilibrium condition
becomes useless.

Recent studies \cite{bia95,adib} have proposed modifications in
the calculation of the entropy that, although irrelevant for large
systems, make significant differences for small systems.
Ref.\cite{bia95} suggests dynamical corrections to the Boltzmann
principle (Eq.(\ref{entrop1})). On the other hand, Ref.\cite{adib}
argues that the entropy in Eqs.(\ref{tempdef}) and
(\ref{entrop1},) should be replaced by
\begin{eqnarray}
S_{\Phi}=k_B\ln \Phi (E),\label{nentro}
\end{eqnarray}
where $n(E)=\frac{d \Phi(E)}{d E}$. It has been shown that
Eqs.(\ref{entrop1}) and (\ref{nentro}) lead to identical results
in the thermodynamic limit, but not for small systems, where
(\ref{nentro}) is still able to describe well the results of
numerical simulations.

For the present model we have $\Phi_{o}(E)\propto E$,
$\Phi_{NS}(E)\propto E^2$ and $\Phi_{QS}(E)\propto E^{3/2}$. In
this framework, the equilibrium condition can be obtained equating
the temperatures of the oscillator and the chaotic system,
computed with the modified entropy Eq.(\ref{nentro}). For the NS
we find
\begin{eqnarray}
\frac{\partial \ln\Phi_o(E_o)}{\partial E_o} = \frac{\partial
\ln\Phi_{NS}(E_c)}{\partial E_c} \qquad \to \qquad
\frac{E_o}{E_c}=\frac{1}{2}
\end{eqnarray}
and for the QS
\begin{eqnarray}
\frac{\partial \ln\Phi_o(E_o)}{\partial E_o} = \frac{\partial
\ln\Phi_{QS}(E_c)}{\partial E_c} \qquad \to \qquad
\frac{E_o}{E_c}=\frac{2}{3}.
\end{eqnarray}
These results are in complete agreement with those obtained via
the probability densities. Considering that our theoretical
equilibrium energies describe very well the numerical calculations
and the agreement between these energies and the thermal
equilibrium conditions, we can conclude that the temperature
$T^\Phi$ is indeed a good parameter for characterizing the
equilibrium.

As a last remark we note that the initial distribution function,
which is microcanonical only in the chaotic degrees of freedom, is
not expected to evolve to a fully microcanonical distribution over
the entire system. This is because the dynamics of the full system
is probably mixed, not ergodic. However, the agreement of the
energy distributions at long times with the above calculation
suggests that the dynamics is at least `close' to ergodic, in the
sense that typical trajectories explore a large fraction of the
available energy shell.

\section{Conclusions}

Our numerical results show that the coupling of an oscillator to a
low dimensional chaotic system simulates very well some aspects of
a Brownian particle in a harmonic potential in the presence of a
thermal bath. In particular, the average energy of the harmonic
oscillator shows irreversible behavior and tends to an equilibrium
at long times. For short time scales the average dynamics of the
oscillator can be studied by linear response theory, where
irreversibility is seen to result from the temporal decay of the
chaotic correlations. The average motion of the oscillator follows
a Langevin type of equation whose frequency-dependent dissipation
depends only the amplitude, frequency and decay rate of the
chaotic correlations.

The initial energies of the sub-systems play an important role in
the problem. $E_c(0)$ defines the dynamical regime of the chaotic
system, and the ratio $E_o(0)/E_c(0)$ defines if the oscillator
will absorb or dissipate energy to the chaotic system. This is
very similar to the usual thermalization of a Brownian particle,
with $E_c(0)$ playing the role of the average energy $k_BT$ of the
thermal bath. An important difference, however, is that the
average energy of the chaotic system is also affected by the
coupling, which is a direct consequence of its small number of
degrees of freedom.

For long times the average energy of both oscillator and chaotic
system tend to equilibrate. The value of the average energy and
the energy distribution at equilibrium can be calculated assuming
equal probabilities of the available microscopic states of the
full system and weak coupling, so that probabilities over the full
system can be approximated by the product of the probabilities
over each subsystem.

It is interesting to see that both the short time and the long
time analysis allow the determination of initial energies $E_o(0)$
and $E_c(0)$ such that no exchange of energies occur in the
average. For short times it is given by the condition $A=0$,
Eq.(\ref{equil}) for the NS system. For the parameters of Fig.3 it
gives $E_o(0)/E_c(0)\approx 0.25$. For long times, since
$\bar{E_o}=E_T/3$, imposing $\bar{E_o}=E_o(0)$ and
$E_T=E_o(0)+E_c(0)$, we find $E_o(0)/E_c(0)=0.5$. The two
estimates clearly disagree, which means that for intermediate
times non-linear corrections to the linear theory become important
and change the short time tendency of the average energy. In other
words, the short time dynamics is completely determined by the
properties of the isolated chaotic system, whereas the long time
behavior is dictated by the statistical properties of the full
system. The same reasoning applies to the QS. Fig.9 shows $\langle
E_o(t)\rangle$ for $E_c(0)=0.38$ and $E_o(0)=E_c(0)/2=0.19$. This
is the condition for no exchange of energies at long times, but
corresponds to a situation where the oscillator should dissipate
at short times. And that is exactly what happens: the long time
behavior displayed in Fig.9(a) shows that the oscillator's average
energy is indeed approximately equal to its starting energy.
However, for short times, Figs.9(a) and (b), it clearly dissipates
energy, re-absorbing it back latter on. The disagreement between
short and long time equilibrium conditions may also be related to
the low dimensionality of the chaotic system \cite{bia94}.

The low dimensionality is also reflected in the fact that the
oscillator's equilibrium energy is different for each initial set
up, clearly distinguishing the small chaotic environment from an
infinite thermal bath. Besides, the oscillator's motion for a
single realization (single initial condition) exhibits large
fluctuations with respect to the average.

Finally, the temperature defined from the volume entropy
$S_{\Phi}$ \cite{adib} describes very well the equilibrium
condition of our model, since it agrees with the equilibrium
conditions found by the probability densities and, consequently,
with the numerical results. We believe this result contributes to
the question of  `\ldots whether the volume or the area entropy is
the correct starting point of thermostatistics for small systems'
\cite{adib}. However, it is important to remember that, although
it is possible to define such a temperature, there are always
large fluctuations around these equilibrium values because of the
low number of degrees of freedom. Moreover, it is not clear why
the probability densities are given by the density of states and
the temperature by its integral.

We note that the validity of the two assumptions we used here to
describe the long time equilibrium regime are the subject of a
debate that touches on the foundations of statistical mechanics
\cite{huang}. The assumption of equal probabilities {\it a priori}
is related to ergodicity and mixing. Here we have applied this
condition without much justification, especially because the
global system Eq.(\ref{modelo}) is probably not mixing. However,
the results we obtained from this assumption agrees very well with
the numerical calculations. Thus, we believe that this type of
low-dimensional
models can be good testing grounds to the study of these topics.\\


\centerline{Acknowledgements} \noindent This paper was partly
supported by the Brazilian agencies {\bf FAPESP}, under contracts
number 02/04377-7 and 03/12097-7, and {\bf CNPq}. Especial thanks to S.M.P.


\newpage

\begin{figure}
\centering
\includegraphics[clip=true,width=6cm,angle=0]{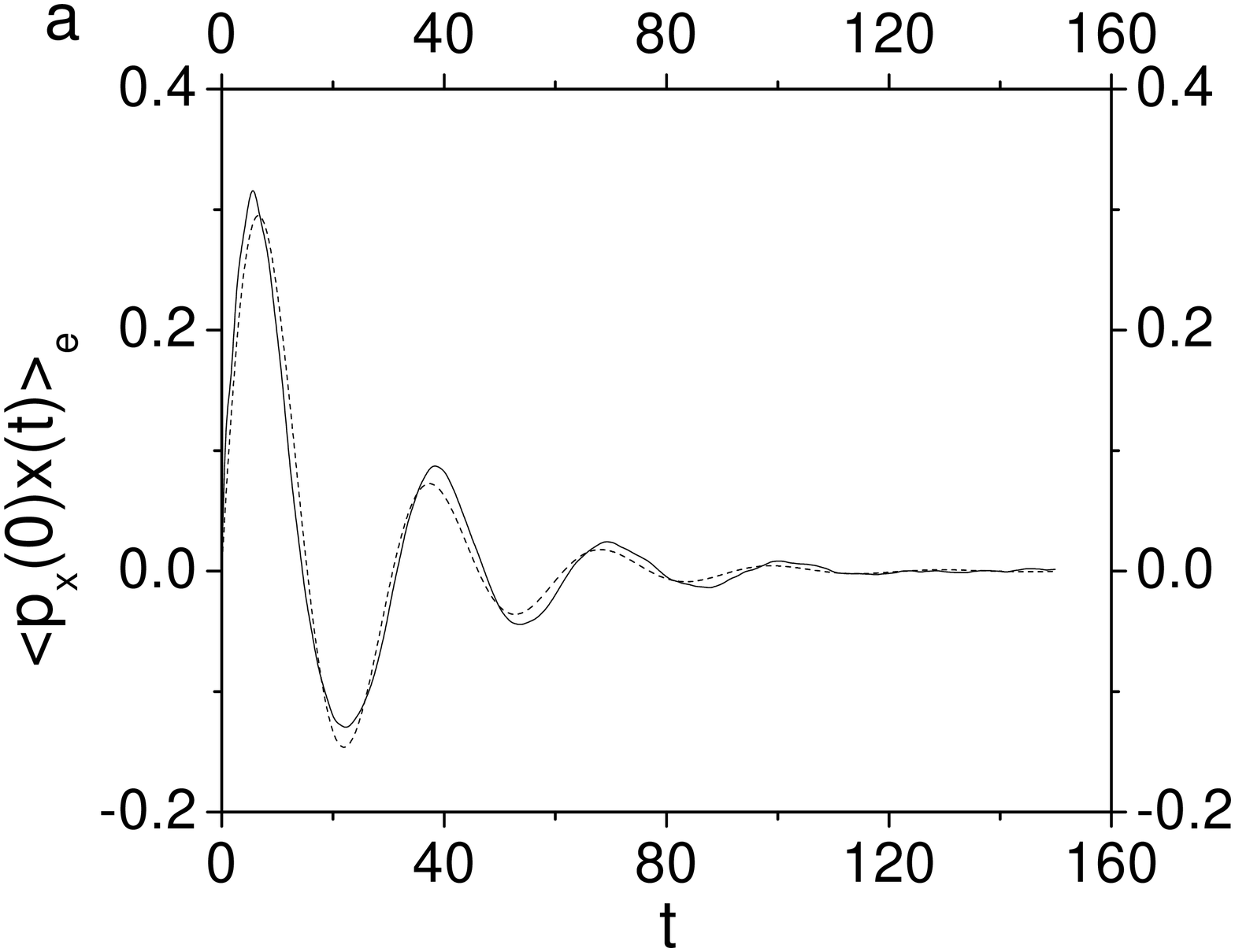}
\includegraphics[clip=true,width=5.7cm,angle=0]{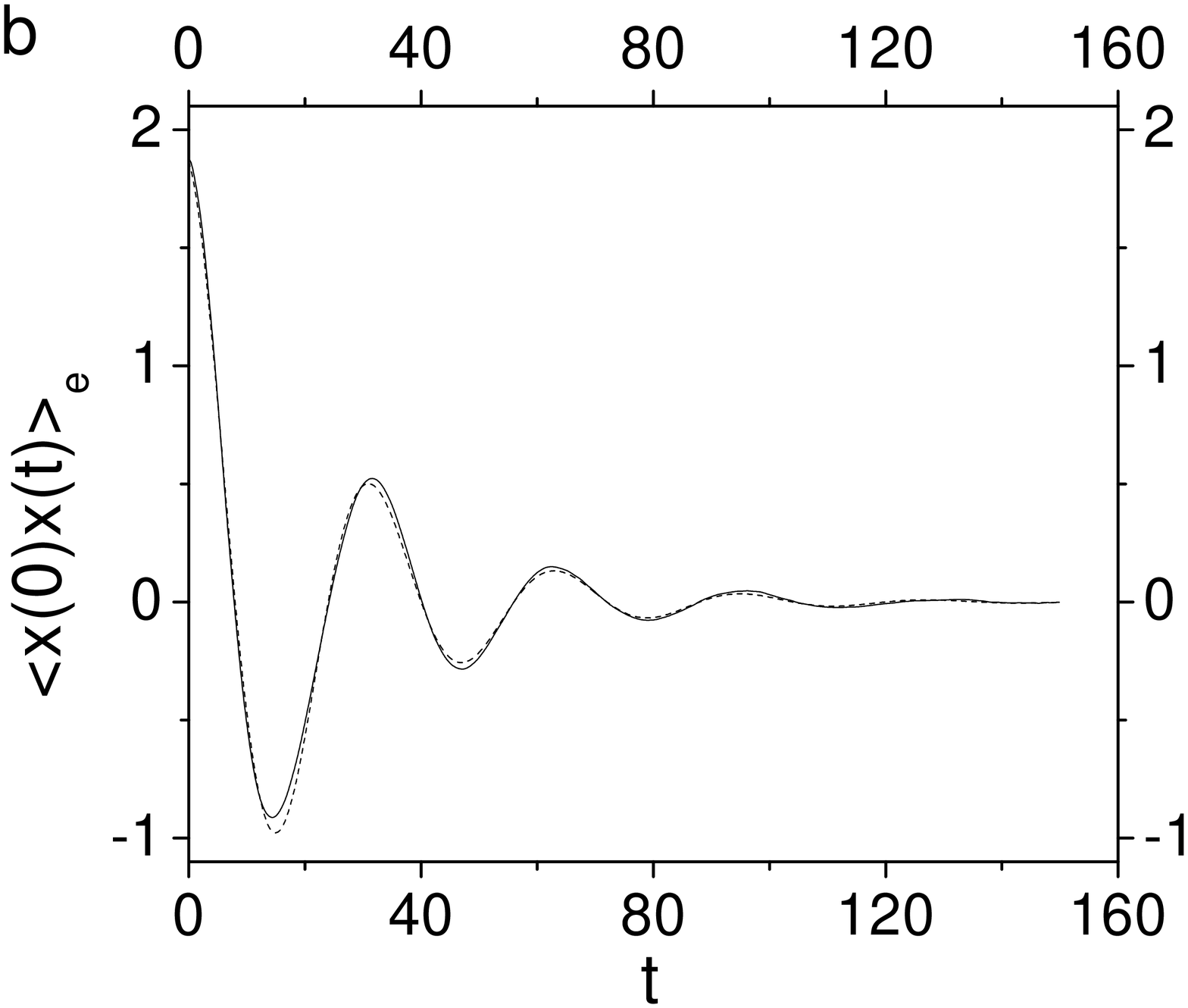}
\label{fig1} \caption{Correlation functions for the NS for
$E_c=0.38$: (a) $\langle p_x (0)x(t)\rangle_e$; (b) $\langle
x(0)x(t)\rangle_e$. The full line shows the numerical results and
the dashed line shows the fitting. The averages were computed
using 35000 initial conditions.}
\end{figure}

\begin{figure}
\centering
\includegraphics[clip=true,width=5.2cm,angle=0]{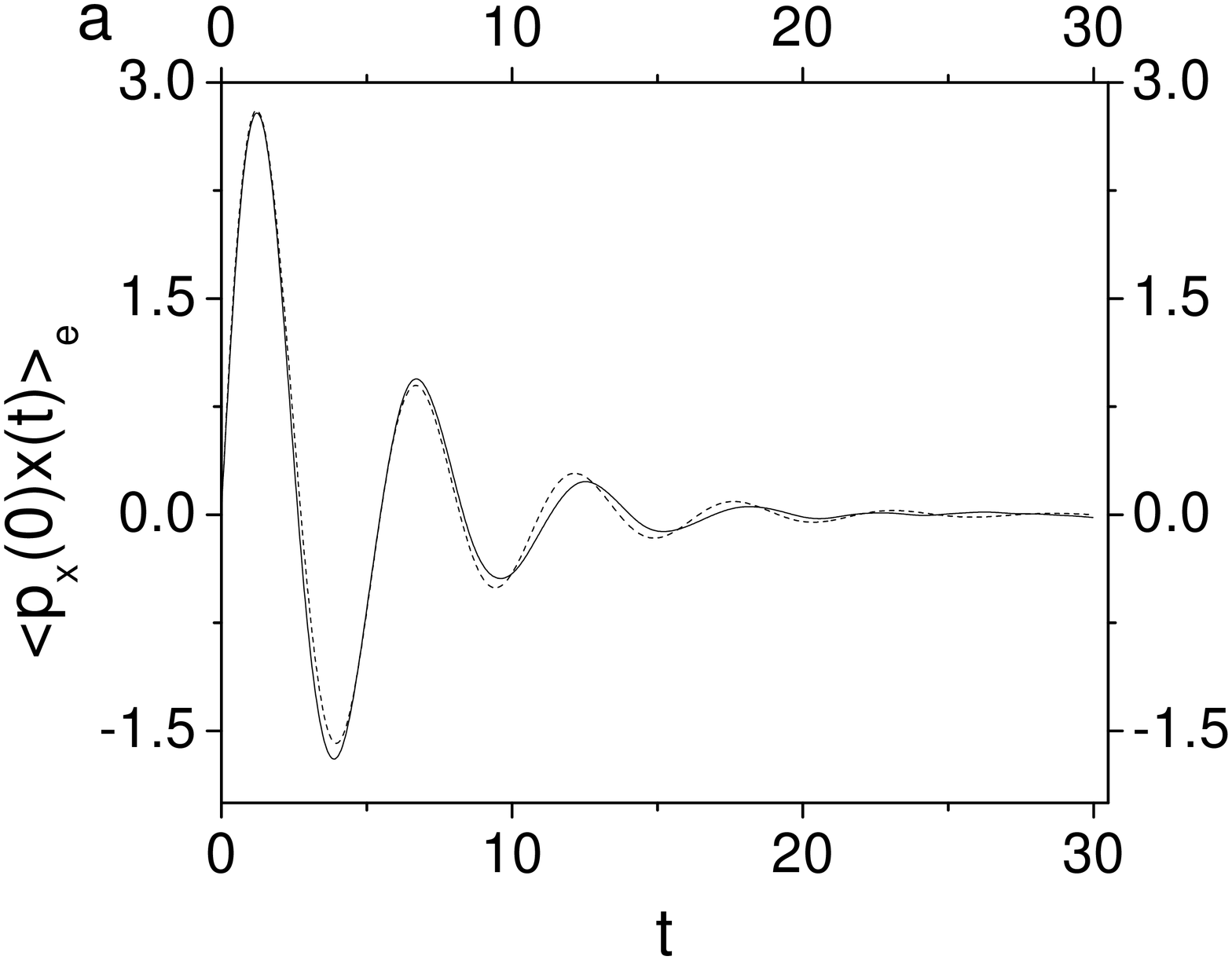}
\includegraphics[clip=true,width=5.2cm,angle=0]{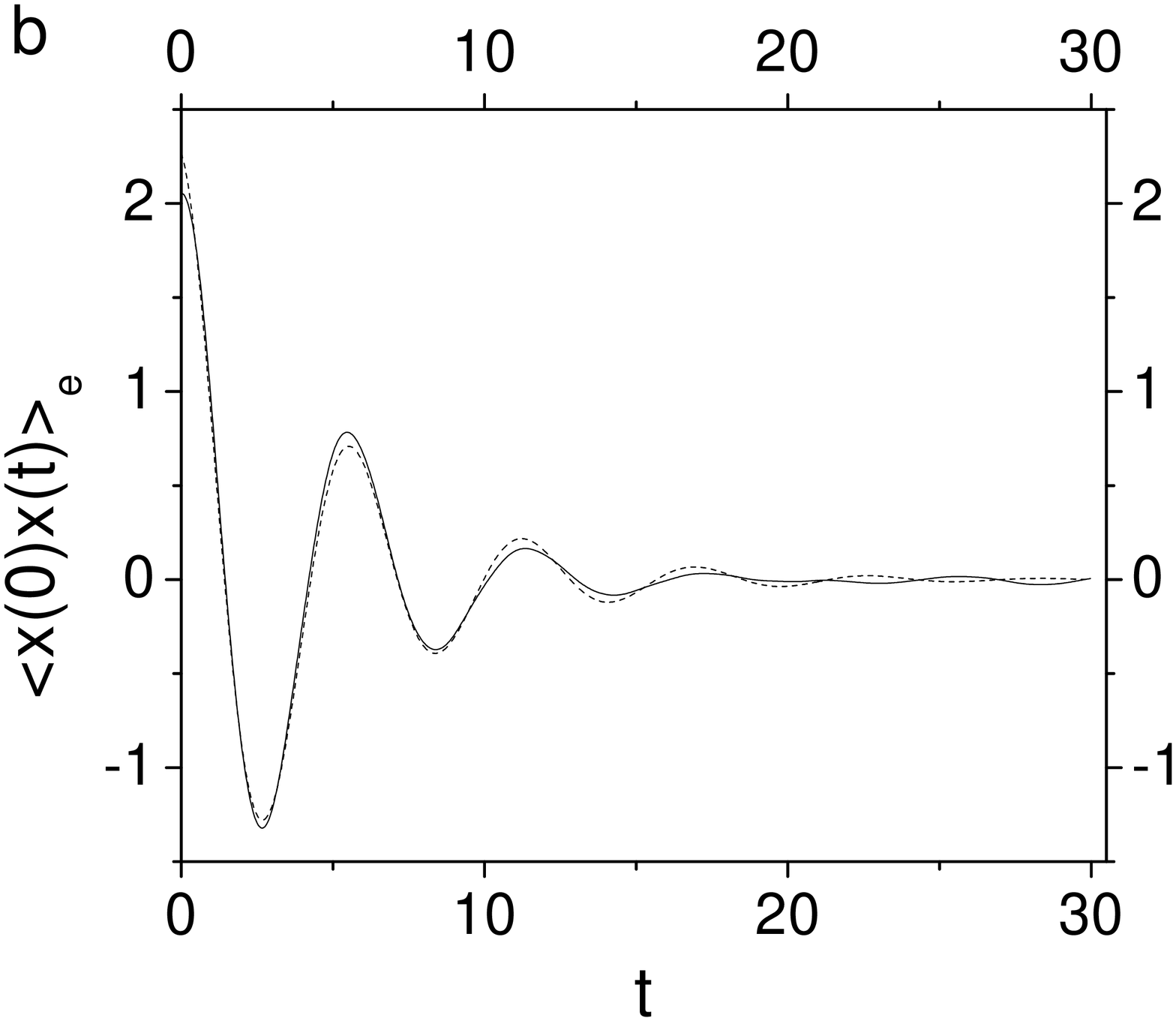}
\includegraphics[clip=true,width=5.2cm,angle=0]{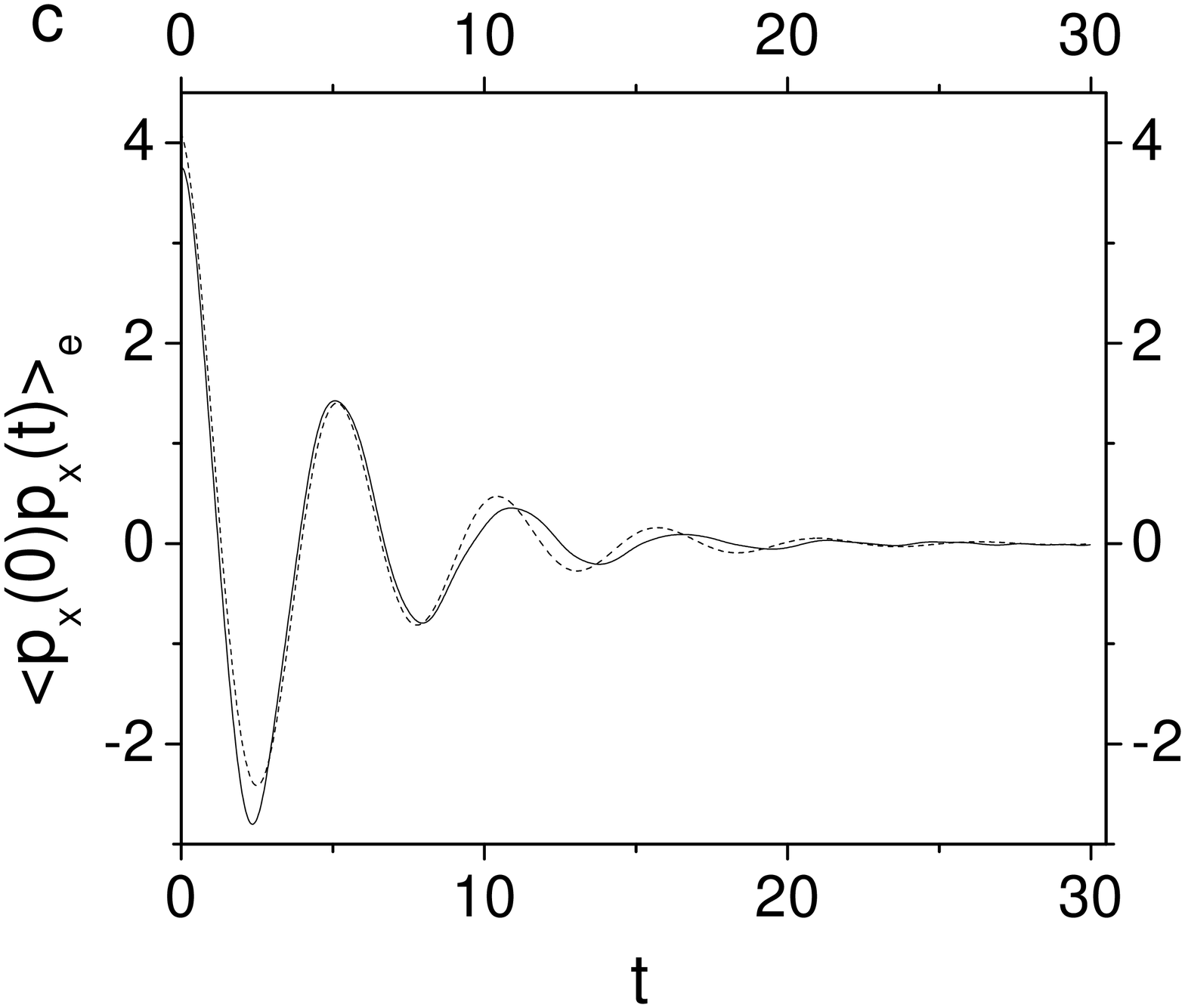}
\label{fig2} \caption{Correlation functions for the QS for
$E_c=5.0$ and $a=0.1$: (a) $\langle p_x (0)x(t)\rangle_e$; (b)
$\langle x(0)x(t)\rangle_e$; and (c) and $\langle
p_x(0)p_x(t)\rangle_e$. The full line shows the numerical results
and the dashed line shows the fitting. The averages were computed
using 30000 initial conditions.}
\end{figure}

\begin{figure}
\centering
\includegraphics[clip=true,width=6cm,angle=0]{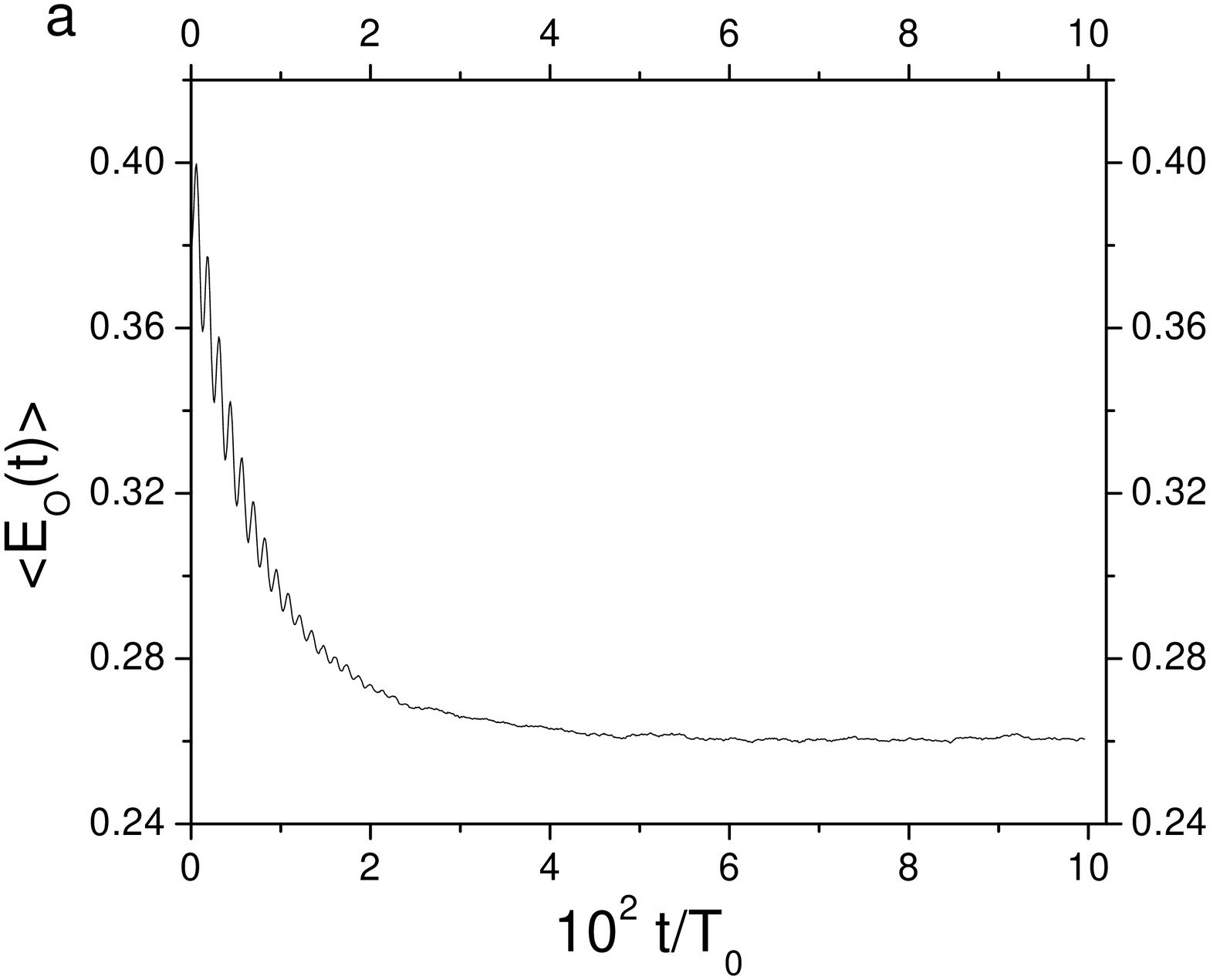}
\includegraphics[clip=true,width=6cm,angle=0]{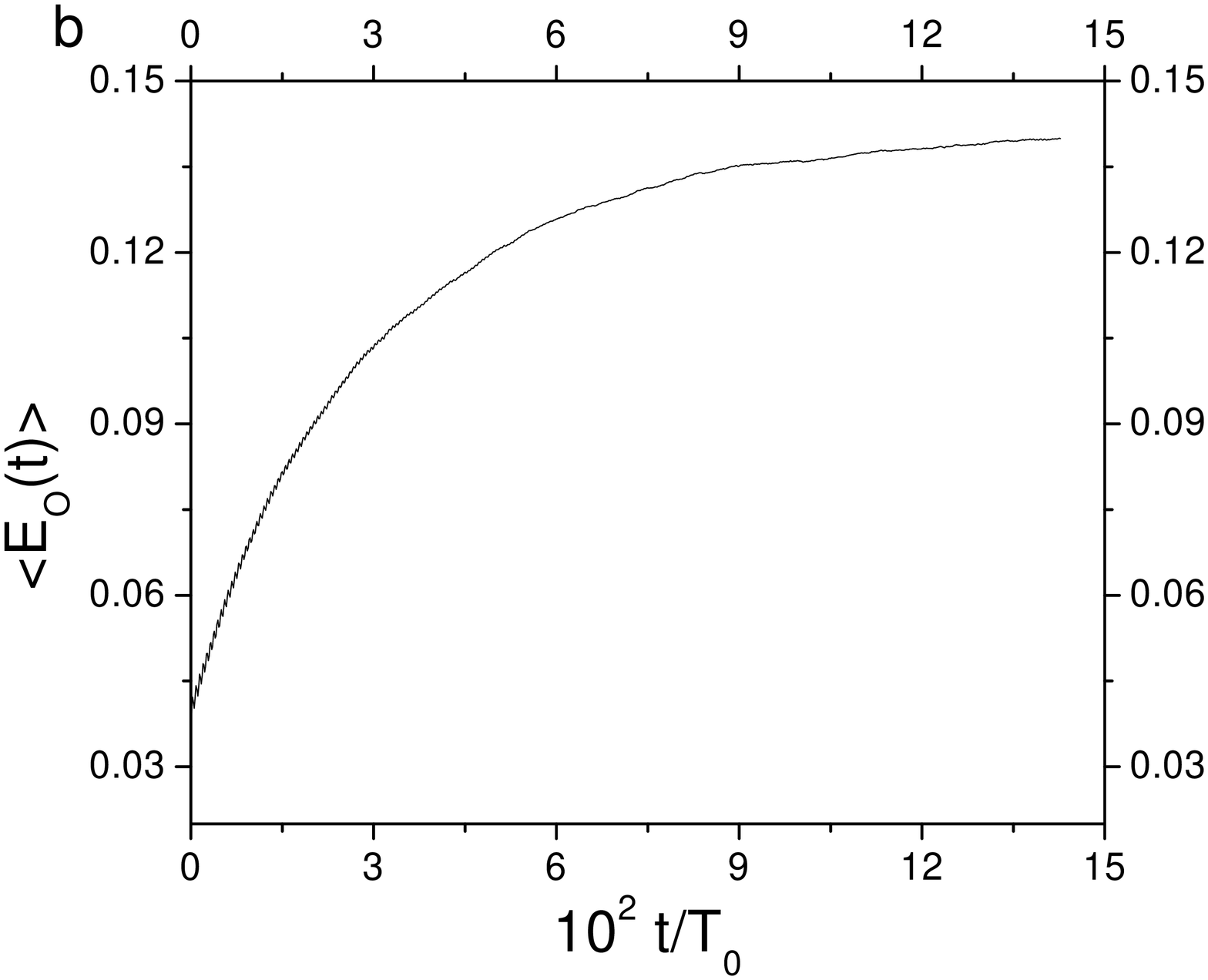}
\label{fig3} \caption{Numerical results for $\langle
E_o(t)\rangle$ with $E_c(0)=0.38$ for NS: (a)$E_o(0)/E_c(0)=1.0$
and (b)$E_o(0)/E_c(0)=0.1$. The oscillator's parameters were set
to $m=200.0$, $\omega_0=0.005$ and the coupling constant to
$\gamma=0.006$. The oscillator's period is $T_0 \approx 1260$. We
used 62480 initial conditions.}
\end{figure}

\begin{figure}
\centering
\includegraphics[clip=true,width=6cm,angle=0]{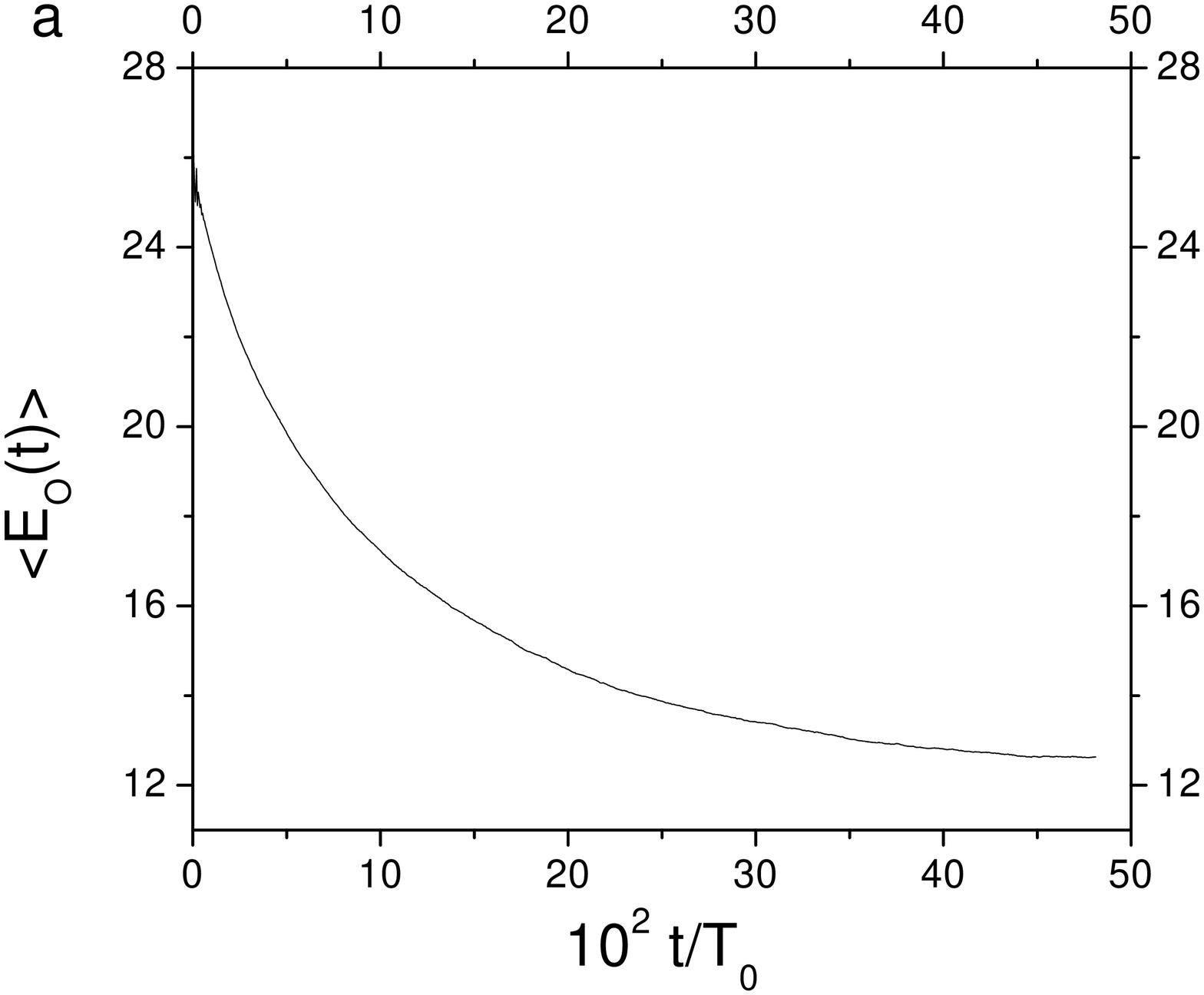}
\includegraphics[clip=true,width=6cm,angle=0]{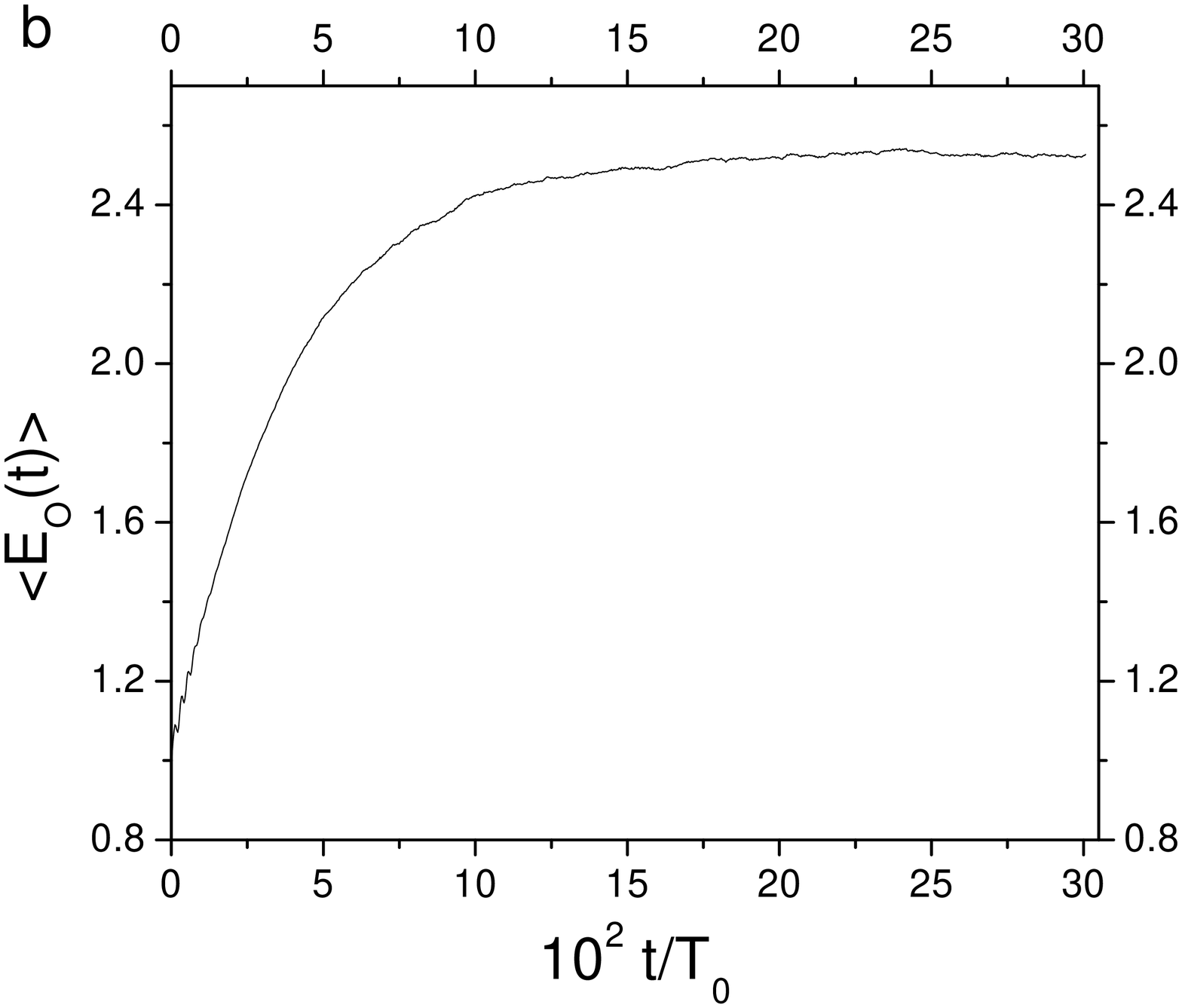}
\label{fig4} \caption{Numerical results for $\langle
  E_o(t)\rangle$ with $E_c(0)=5.0$ and $a=0.1$ for the QS:
(a)$E_o(0)/E_c(0)=5.0$ and (b)$E_o(0)/E_c(0)=0.2$. The
oscillator's parameters were set to $m=10.0$, $\omega_0=0.01$ and
the coupling constant to $\gamma=0.01$. The oscillator's period is
$T_0 \approx 628$. We used 40000 initial conditions.}
\end{figure}

\begin{figure}
\centering
\includegraphics[clip=true,width=5cm,angle=0]{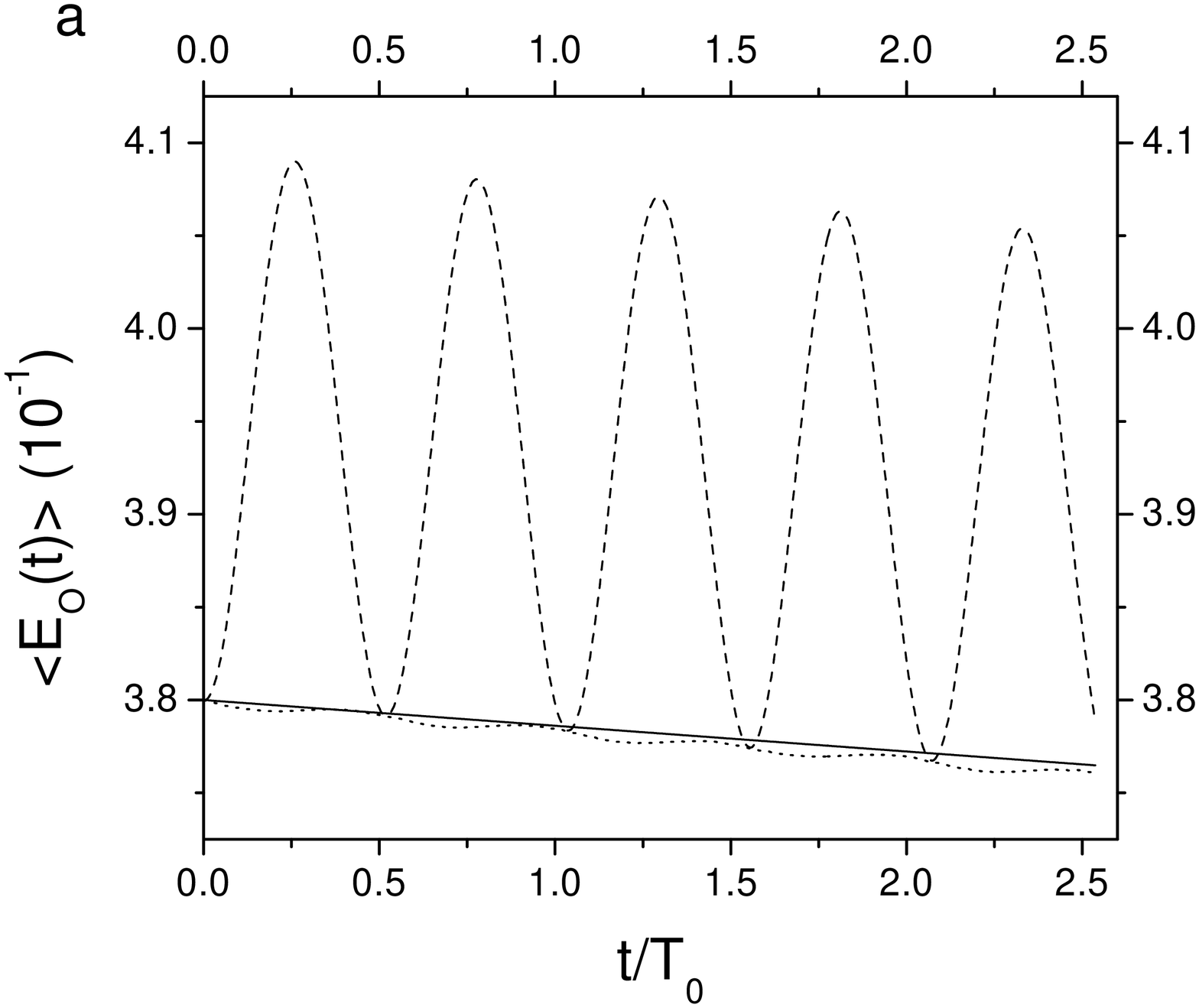}
\includegraphics[clip=true,width=5cm,angle=0]{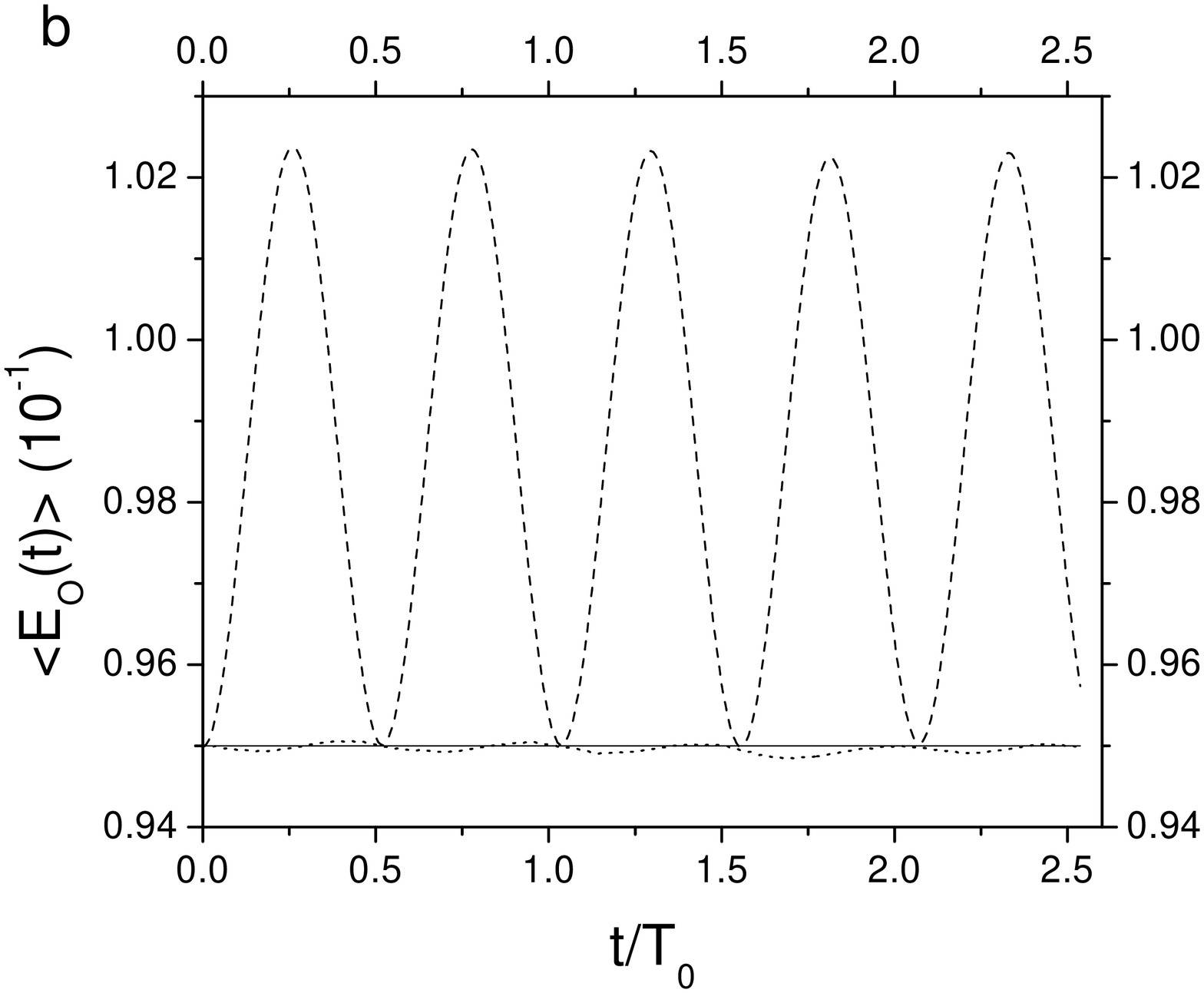}
\includegraphics[clip=true,width=5cm,angle=0]{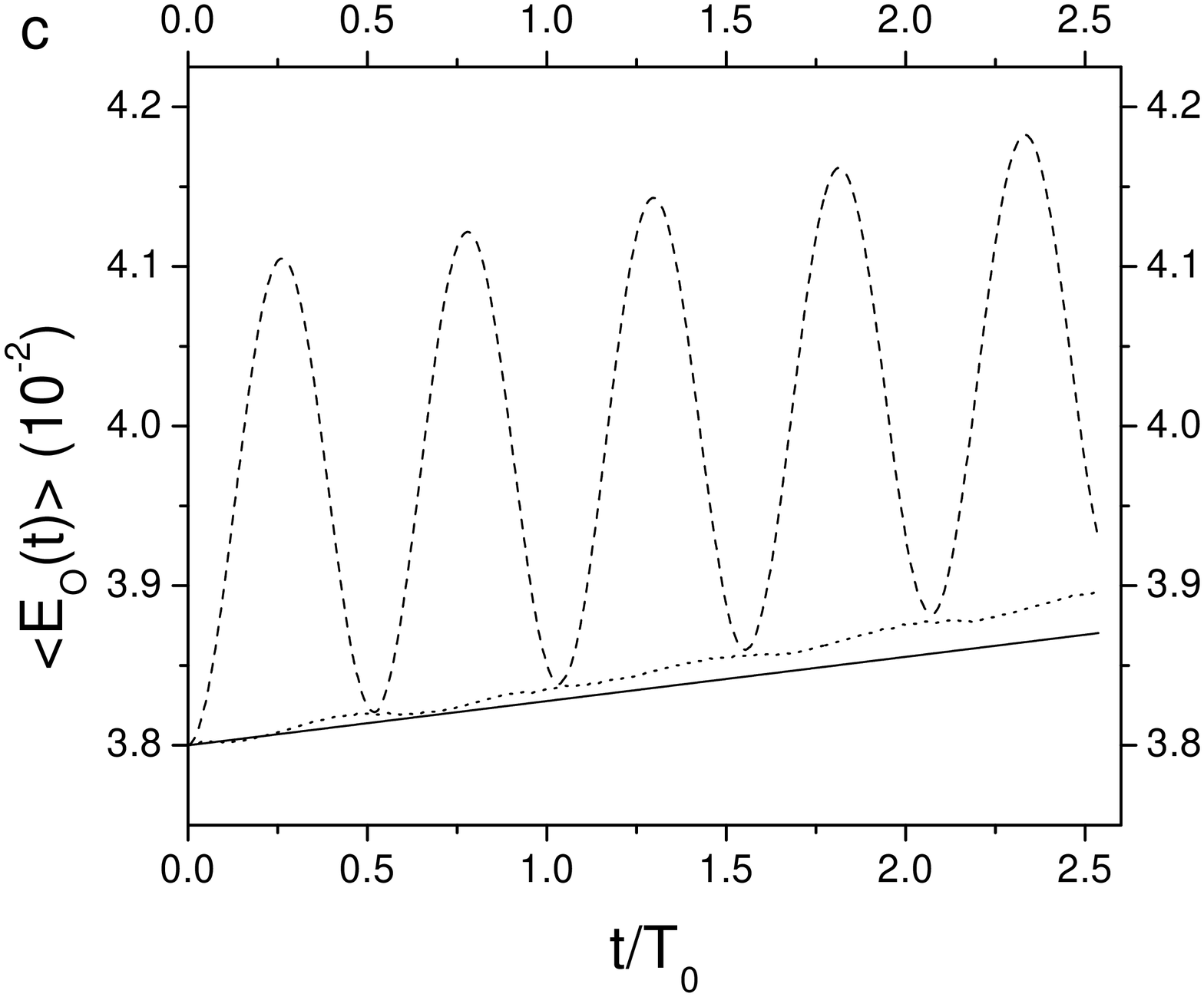}
\label{fig5} \caption{Average oscillator energy at short times
  with the NS as chaotic system. The dashed line shows $\langle
  E_o(t)\rangle$ and the doted line shows $\langle
  E_{or}(t)\rangle$, both obtained numerically. The full line
  shows Eq.(\ref{eoft}) without $f(t)$. (a)
  $E_o(0)/E_c(0)=1.0$, (b) $E_o(0)/E_c(0)=0.25$ and (c)
  $E_o(0)/E_c(0)=0.1$. The oscillator's parameters, coupling
  constant and number of initial conditions are the same as in Fig.3.}
\end{figure}

\begin{figure}
\centering
\includegraphics[clip=true,width=6cm,angle=0]{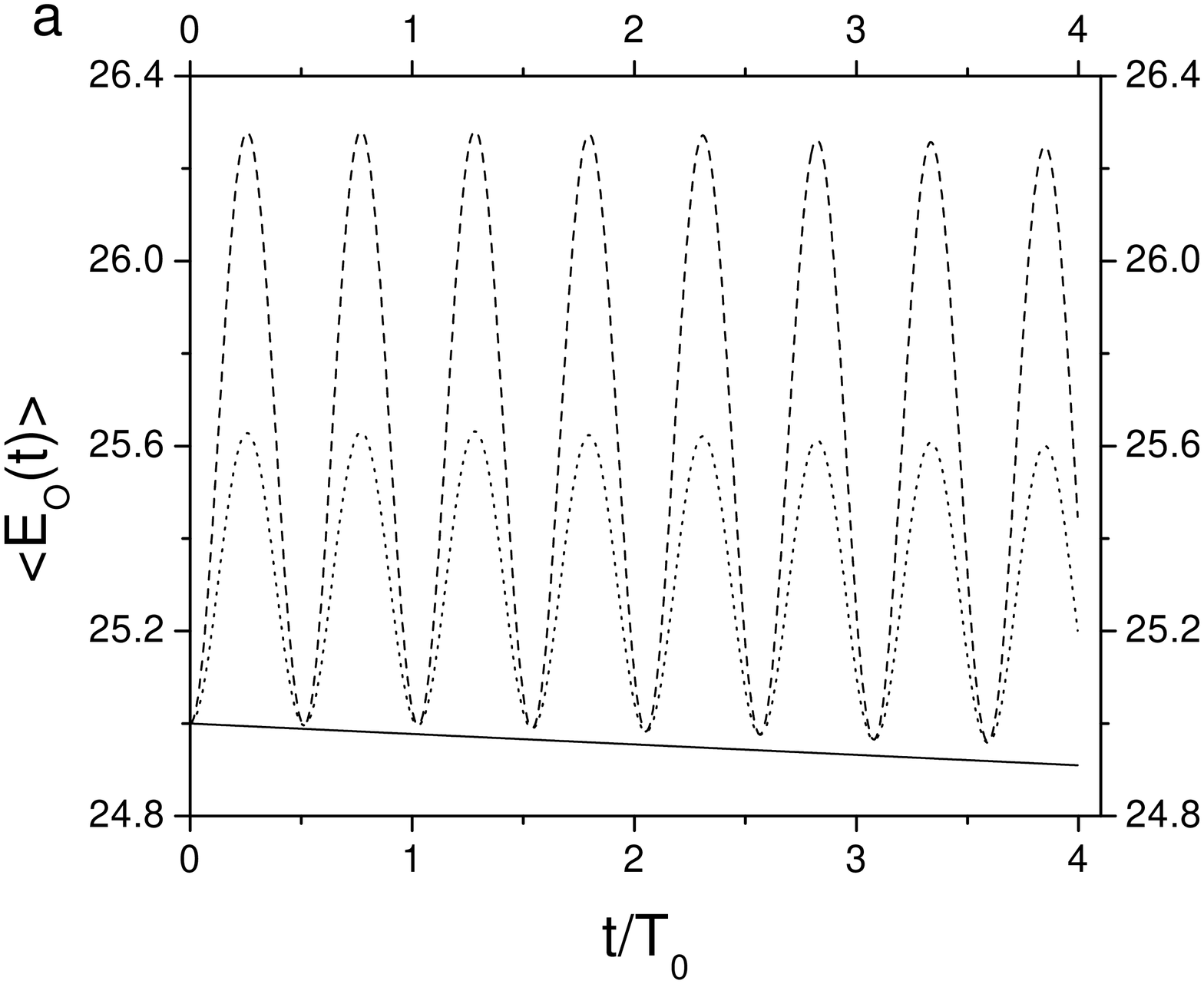}
\includegraphics[clip=true,width=6cm,angle=0]{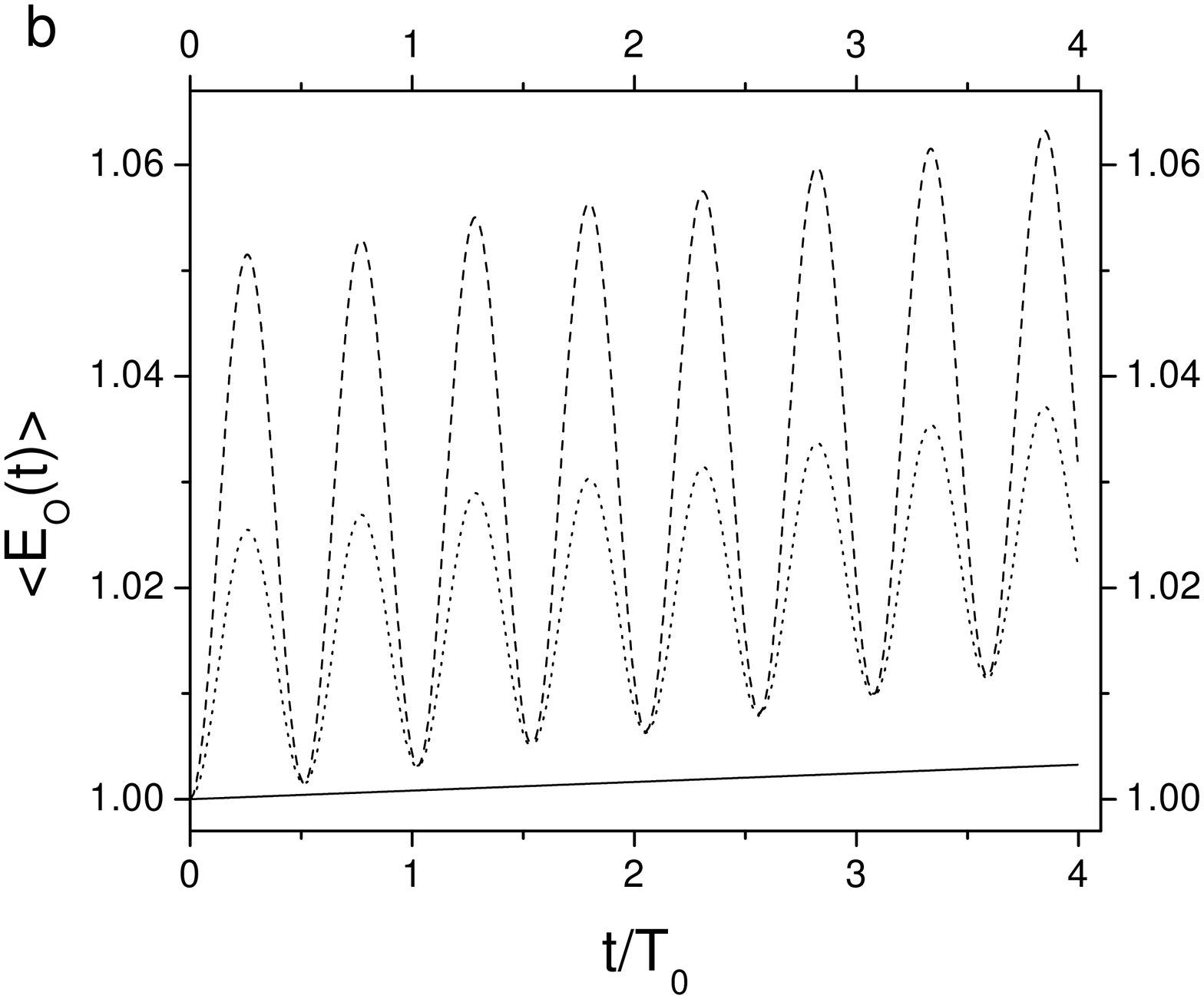}
\label{fig6} \caption{Average oscillator energy at short times
  with the QS as chaotic system. The dashed line shows $\langle
  E_o(t)\rangle$ and the doted line shows $\langle
  E_{or}(t)\rangle$, both obtained numerically. The full line
  shows Eq.(\ref{eneravq}) without $f'(t)$.
  (a) $E_o(0)/E_c(0)=5.0$ and (b) $E_o(0)/E_c(0)=0.2$. The
  oscillator's parameters, coupling constant and number of
  initial conditions are the same as in Fig.4.}
\end{figure}

\begin{figure}
\centering
\includegraphics[clip=true,width=6cm,angle=0]{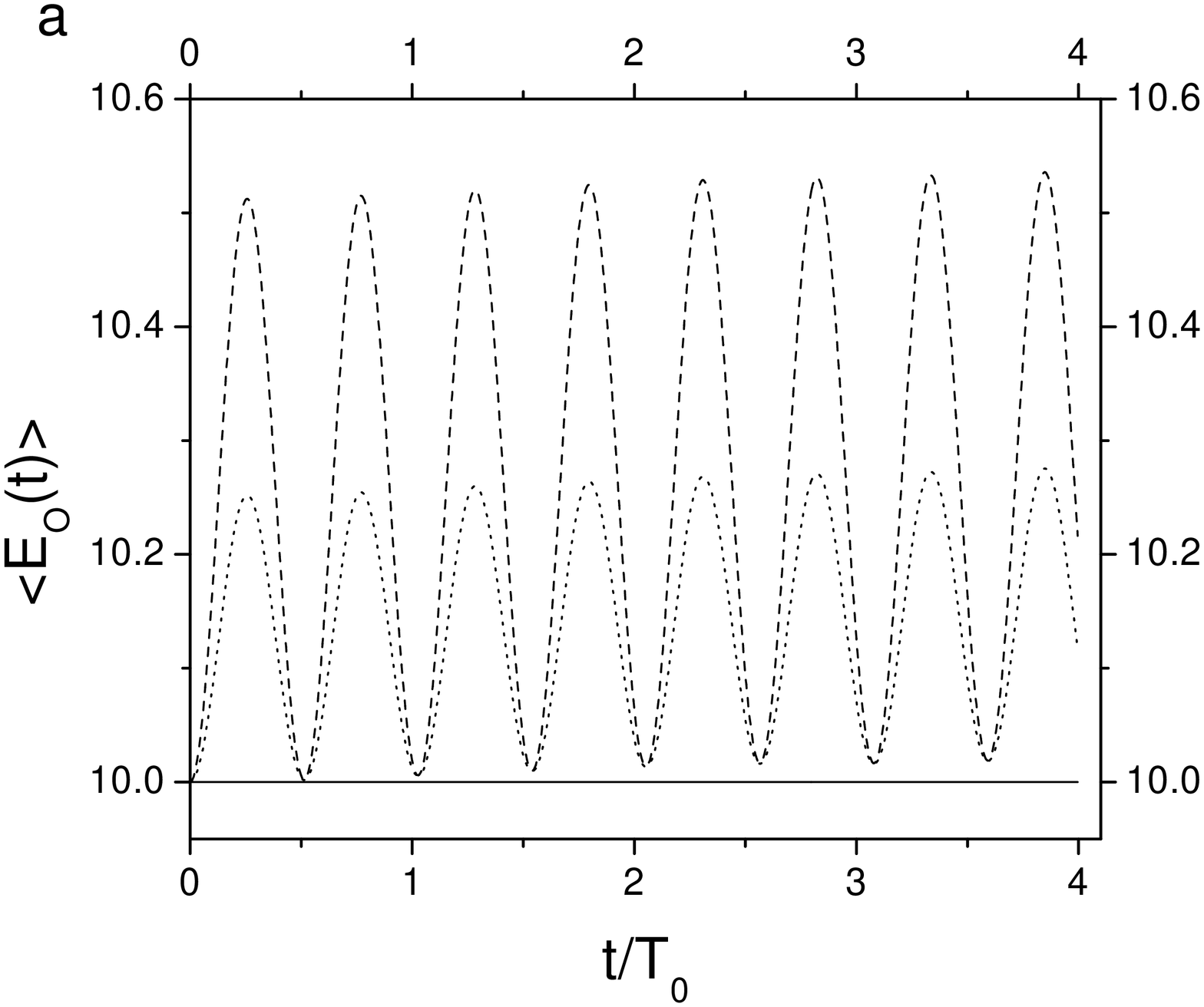}
\includegraphics[clip=true,width=6cm,angle=0]{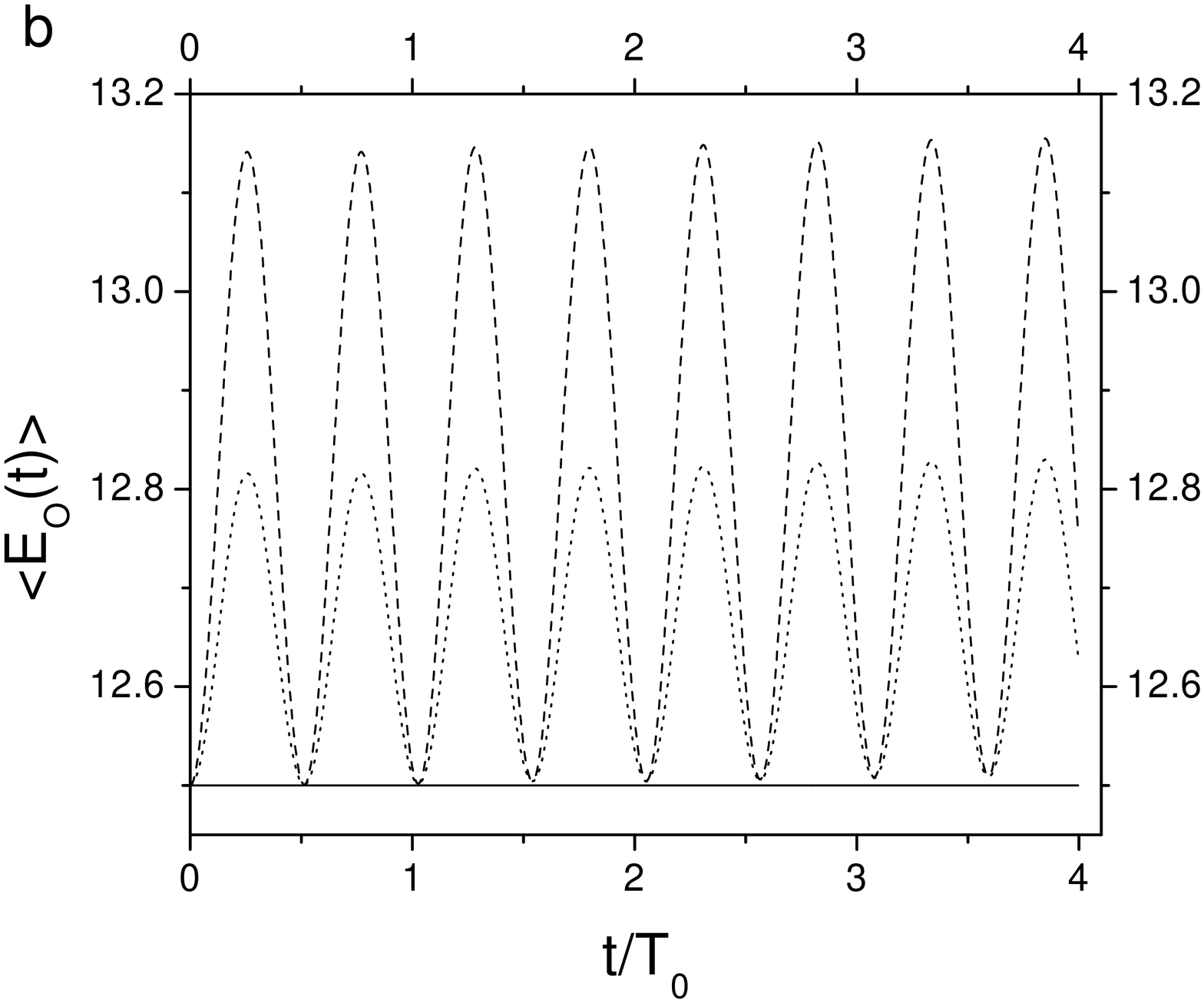}
\includegraphics[clip=true,width=6cm,angle=0]{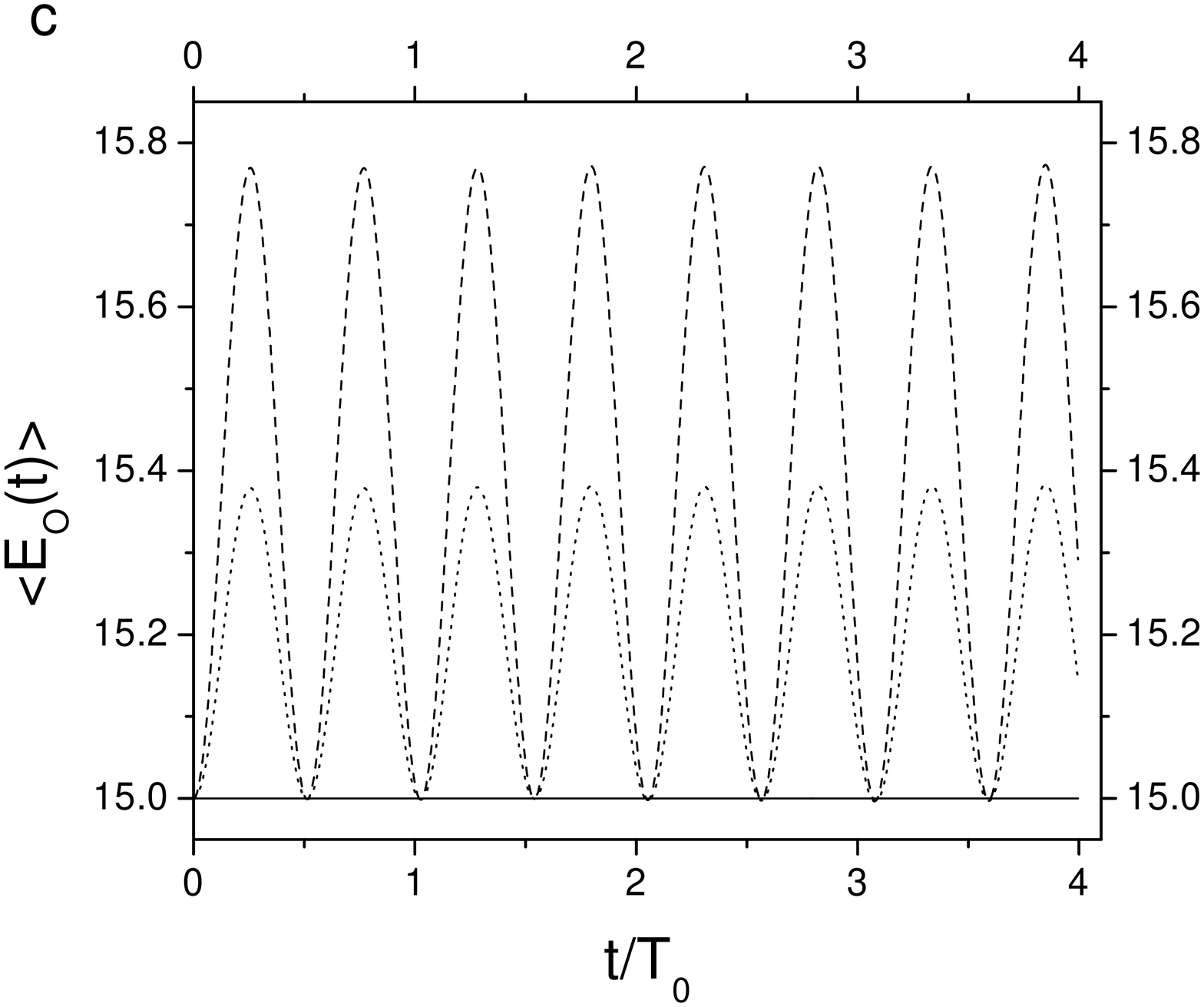}
\includegraphics[clip=true,width=6cm,angle=0]{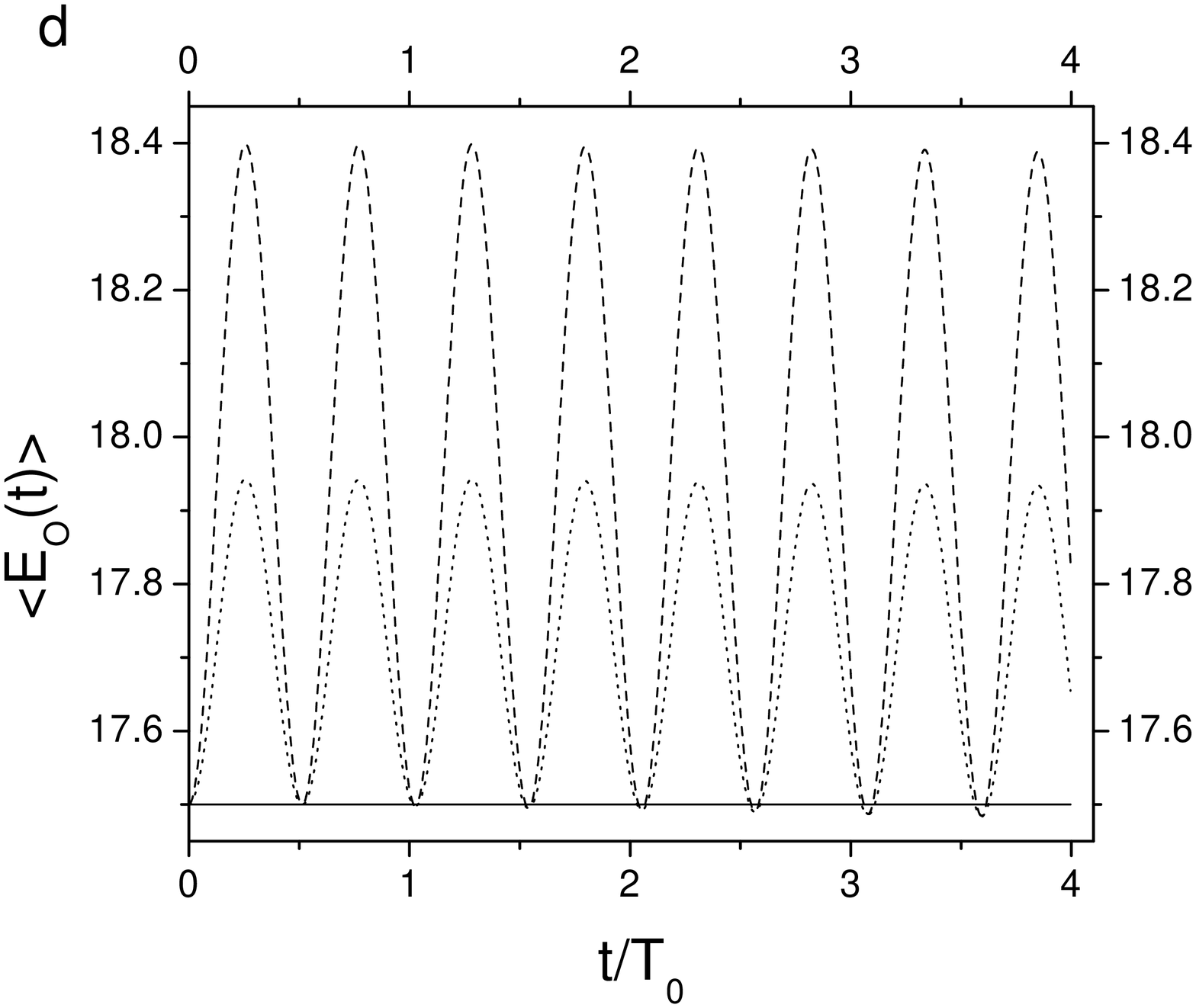}
\label{fig7} \caption{Average oscillator energy at short times
  with the QS as chaotic system. The dashed line shows $\langle
  E_o(t)\rangle$ and the doted line shows $\langle
  E_{or}(t)\rangle$, both obtained numerically. The full line
  corresponds to $E_o(0)$. (a) $E_o(0)/E_c(0)=2.0$,
  (b) $E_o(0)/E_c(0)=2.5$, (c)$E_o(0)/E_c(0)=3.0$
  and $E_o(0)/E_c(0)=3.5$. The oscillator's parameters, coupling
  constant and number of initial conditions are the same as in Fig.4.}
\end{figure}

\begin{figure}
\centering
\includegraphics[clip=true,width=6cm,angle=0]{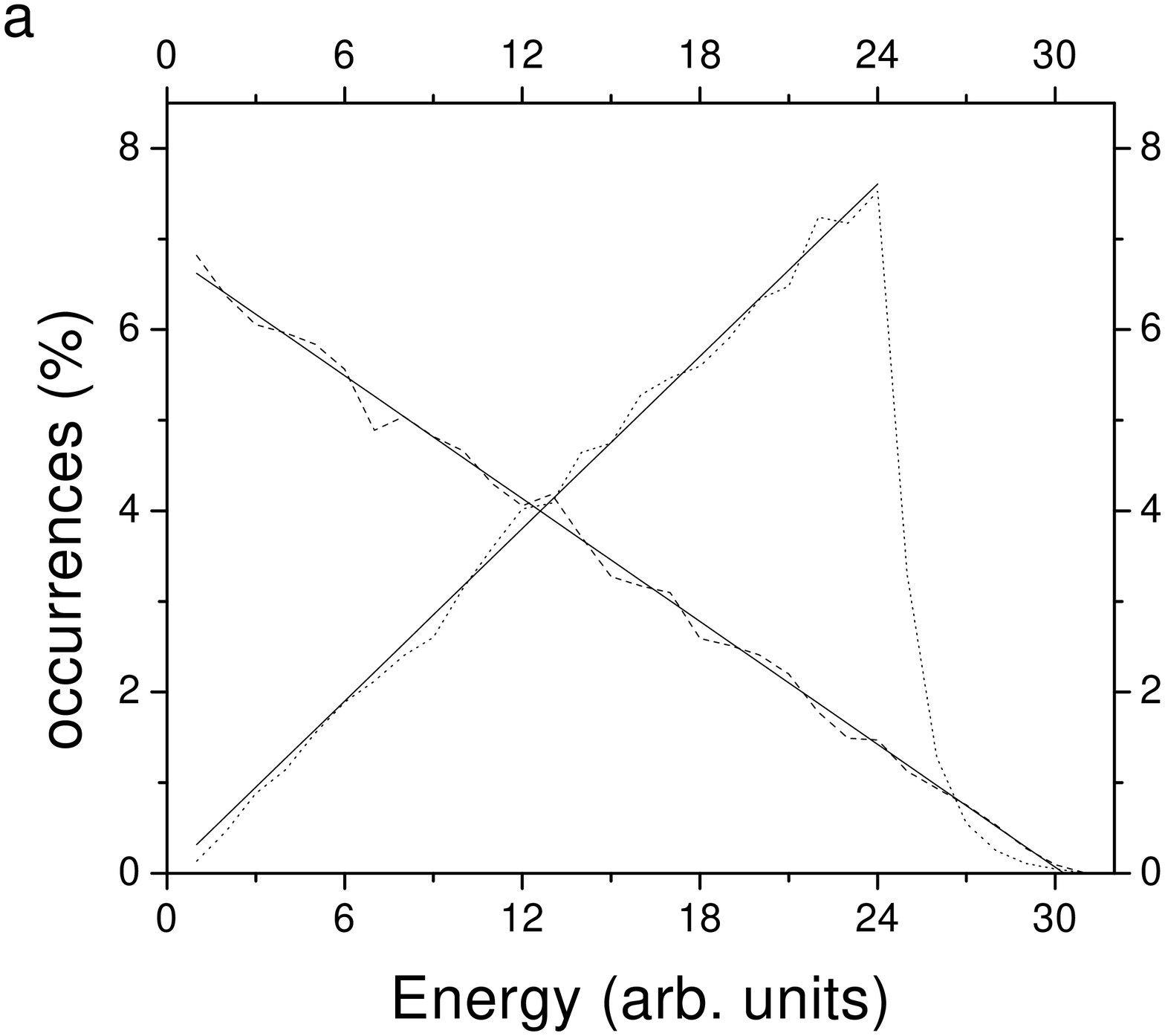}
\includegraphics[clip=true,width=6cm,angle=0]{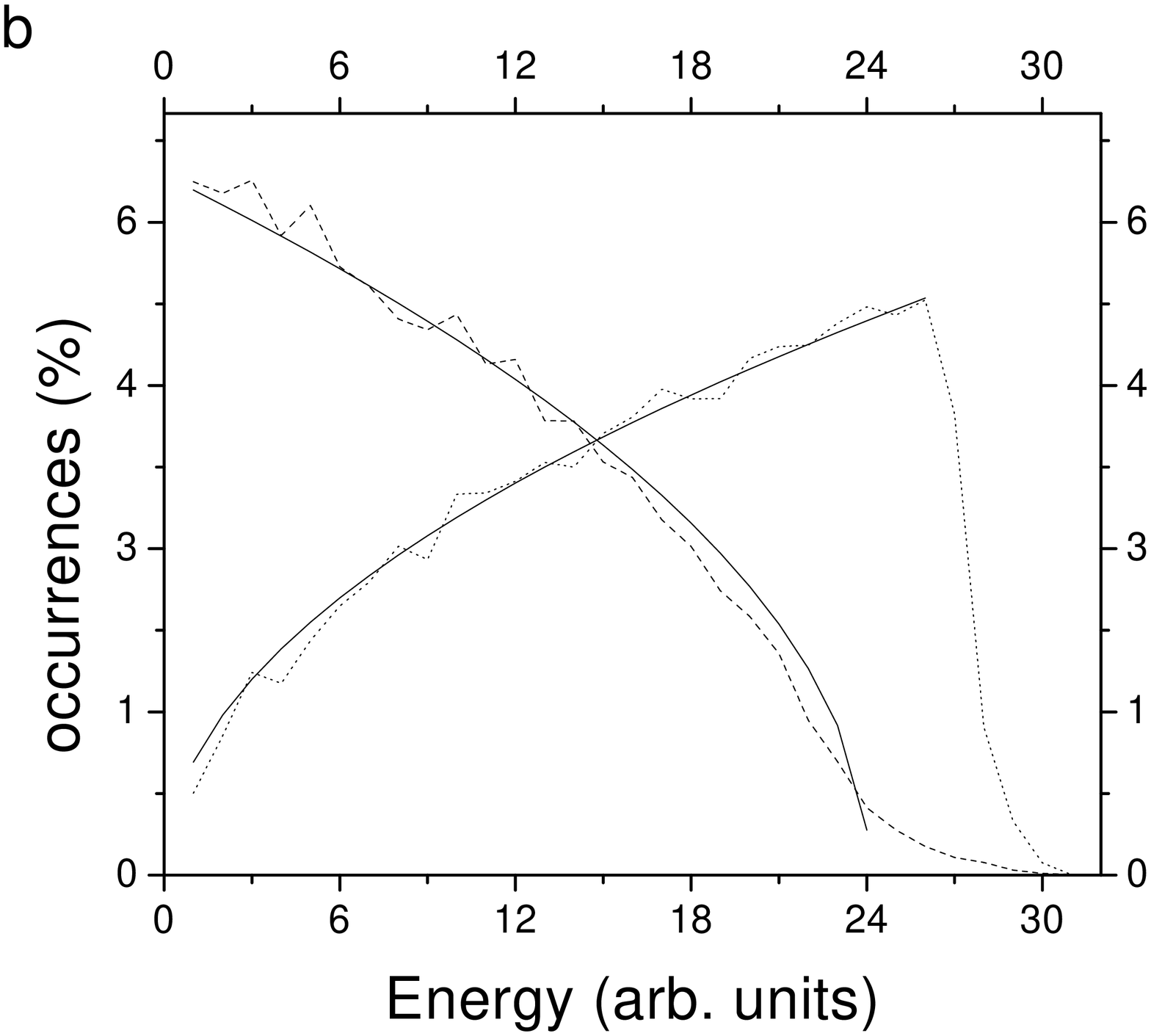}
\label{fig8} \caption{(a)Distribution of energies of the ensemble
of 30000 trajectories at the time $t=8 × 10^{5}$, for the
parameters of Fig. 3a. The full line shows to linear fitting, the
dashed line shows $E_o$ and dotted line shows $E_c$; (b)
Distribution of energies at the time $t=16 × 10^{5}$, for the
parameters of Fig. 4b and 30000 trajectories. The full line
corresponds to square root fitting, the dashed line to $E_o$ and
the dotted line to $E_c$. The energy is shown in units of
$E_T/30$.}
\end{figure}

\begin{figure}
\centering
\includegraphics[clip=true,width=6cm,angle=0]{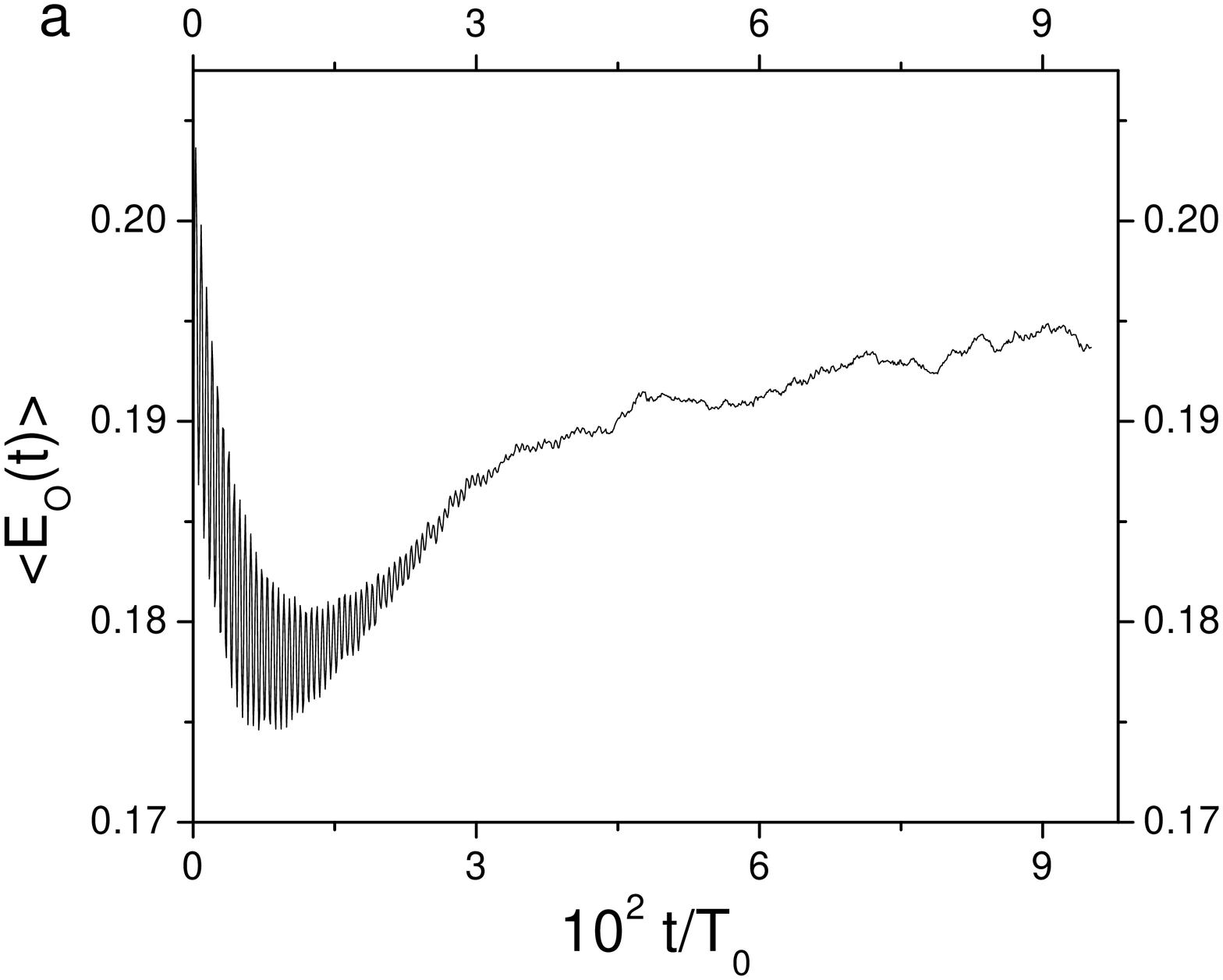}
\includegraphics[clip=true,width=6cm,angle=0]{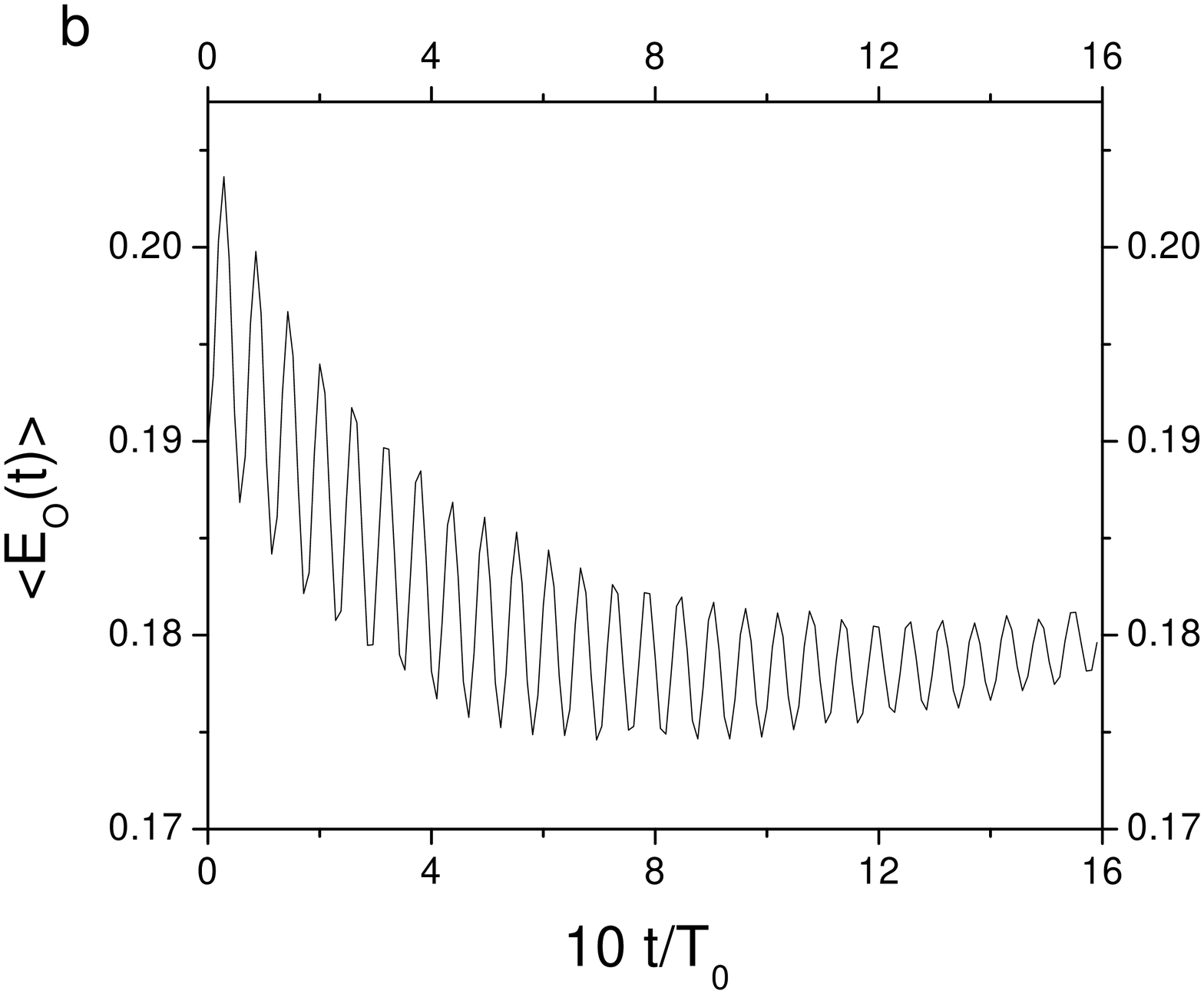}
\label{fig9} \caption{(a) $\langle E_o(t)\rangle$ for
$E_c(0)=0.38$ and $E_o(0)=E_c(0)/2=0.19$. (b) Magnification of the
time interval from 0 to $2 × 10^5$ showing the initial
dissipation of energy.}
\end{figure}

\end{document}